\documentclass[twocolumn,amsmath,amssymb,superscriptaddress,pra,aps]{revtex4}

\usepackage{graphicx} % graphics
\usepackage{picture} % insert figures as tex files
\usepackage{epstopdf} % insert eps files
\usepackage{pdfpages} % insert pdf files
\usepackage{microtype} % fixes spaces
\usepackage{tikz}
\usepackage{xspace}
\usepackage{hyperref} % in order to insert links

\begin{document}

\title{CONAN -- the cruncher of local exchange coefficients for strongly 
interacting confined systems in one dimension}

\date{\today}

\author{N.~J.~S. Loft}
\affiliation{Department of Physics and Astronomy, Aarhus University, DK-8000 Aarhus C, Denmark}
\author{L.~B. Kristensen}
\affiliation{Department of Physics and Astronomy, Aarhus University, DK-8000 Aarhus C, Denmark}
\author{A.~E. Thomsen}
\affiliation{Department of Physics and Astronomy, Aarhus University, DK-8000 Aarhus C, Denmark}
\affiliation{{CP}$^{ \bf 3}${-Origins} \& the Danish Institute for Advanced Study {\rm{Danish IAS}}, University of Southern Denmark, DK-5230 Odense M, Denmark.}
\author{A.~G. Volosniev}
\affiliation{Institut f{\"u}r Kernphysik, Technische Universit{\"a}t Darmstadt, 64289 Darmstadt, Germany}
\affiliation{Department of Physics and Astronomy, Aarhus University, DK-8000 Aarhus C, Denmark}
\author{N.~T. Zinner}
\affiliation{Department of Physics and Astronomy, Aarhus University, DK-8000 Aarhus C, Denmark}

\begin{abstract}
  We consider a one-dimensional system of particles with strong
  zero-range interactions.  This system can be mapped onto a spin
  chain of the Heisenberg type with exchange coefficients that depend
  on the external trap. In this paper, we present an algorithm that
  can be used to compute these exchange coefficients. We introduce an
  open source code CONAN (Coefficients of One-dimensional N-Atom
  Networks) which is based on this algorithm. CONAN works with
  arbitrary external potentials and we have tested its reliability for
  system sizes up to around 35 particles. As illustrative examples, we
  consider a harmonic trap and a box trap with a superimposed
  asymmetric tilted potential. For these examples, the computation
  time typically scales with the number of particles as $O(N^{3.5 \pm
    0.4})$. Computation times are around 10 seconds for $N=10$
  particles and less than 10 minutes for $N=20$ particles.
\end{abstract}
%\pacs{67.85.-d,03.65.Ge,05.30.Fk}
\maketitle

%%%%%%%%%%%%%%%%%%%%%%%%% BEGIN TEXT %%%%%%%%%%%%%%%%%%%%%%%%%%%%%

\section{Introduction} 
The study of strongly interacting systems can be motivated by the 
realization that coherent quantum phenomena, such as 
high-temperature superconductivity, Helium superfluidity, and
quantum magnetism, occur in systems where the particle
interactions are intrinsically strong. In one-dimensional (1D)
systems, examples are strong Coulomb interactions in quantum nanowires 
and in nanotubes \cite{nazarov2009,deshpande2010,bezryadin2013,altomare2013}, as well as linear compounds with strong exchange 
interactions that provide realizations of various spin chains \cite{blundell2004,mourigal2013,sahling2015}.
To treat these 1D systems, some popular
approaches include bosonization and Luttinger liquid theory \cite{giamarchi2003}, 
the Bethe ansatz \cite{takahashi1999}, and, most recently, the density matrix renormalization group 
\cite{ulli2005,ulli2011} invented by Steven White.

It has now become possible to build 
setups with cold atomic gases that can provide quantum simulation 
of these low-dimensional quantum few- and many-body systems \cite{lewenstein2007,bloch2008,esslinger2010,baranov2012,zinner2013}.
In particular, the Tonks-Girardeau gas \cite{tonks1936,girardeau1960} consisting of 
strongly interacting bosons in 1D was realized in 
experiments \cite{olshanii1998,paredes2004,kinoshita2004,kinoshita2006,haller2009}.
Luttinger-liquid behavior in interacting bosonic systems was also observed \cite{haller2010}.
More recently, strongly interacting fermionic systems were realized
in 1D \cite{pagano2014}, also in the limit of just a few particles 
\cite{serwane2011,zurn2012,wenz2013,murmann2015a}. In the 
latter setup it was shown that one can experimentally 
access strongly interacting few-body systems in the regime 
where they realize a Heisenberg spin chain \cite{murmann2015b}.
These developments were a motivation for our work during
the last few years which resulted in the method presented below.

Before we proceed with our presentation, we overview 
some previous studies relevant to the present paper. In 1960 
Girardeau  \cite{girardeau1960} demonstrated the 
connection between strongly interacting bosons and
spin-polarized/spinless fermions. 
For instance, it turns out that the ground state of a bosonic system can be obtained by taking the 
absolute value of the corresponding fermionic wave function. 
This connection was subsequently generalized in various ways. In 2004, within the context of atomic systems, 
Girardeau and Olshanii provided the so-called Fermi-Bose
mapping strategy to describe two-component Fermi and Bose systems 
\cite{girardeau2004}. In 2006 Girardeau connected these observations
to exchange interactions, thus, relating the mapping to spin 
models \cite{girardeau2006}. In the strongly interacting 
regime a formal mapping that produces eigenstates of the 
total spin and the spin projection along one direction was
proposed not long after \cite{deuretz2008,guan2009}.
Note that this connection was also established
within condensed-matter physics, where in 1990
Ogata and Shiba \cite{ogata1990} used the Bethe 
ansatz solution to show how the 1D Hubbard model 
becomes a spin model for strong interactions. The spin models that were derived
have exchange coefficients that are site-independent because the starting point
was usually either homogeneous 1D space or a lattice model with 
some discrete translation invariance. 

It was not until around 2013, when scientists realized that once you 
deviate from the homogeneous case, the spin models and particularly their 
exchange coefficients should be site-dependent, as they need to 
reflect the geometric landscape of the confinement potential. This was initially identified as a short-coming of the Fermi-Bose
mapping in the case of harmonic confinement \cite{gharashi2013,lindgren2014}. 
It was then realized that this does not spell the end 
of the Fermi-Bose mapping and effective spin model Hamiltonians
\cite{volosniev2014,deuretzbacher2014, volosniev2015}, but it does mean that a 
different model that takes the external confinement into 
account should be used. This model implies that a confined $N$-body
system has a set of distinct 'local exchange coefficients' 
that are completely specified by the external potential.
The first reference to provide an explicit formula for these
coefficients is Volosniev~{\it et al.}~\cite{volosniev2014}. In this paper  
it was also realized that the exchange coefficients do 
not depend on the system composition (bosons/fermions/mixtures, two- or 
more internal degrees of freedom) as long as the interactions between 
all species are strong and the particle masses are the same. 
This paper was soon followed by several other papers exploring various
approximations to these coefficients \cite{levinsen2015} and 
their applications \cite{cui2014a,cui2014b,yang2015a,yang2015b,loft2015,pietro2015,marchukov2015,yang2016}.

Due to the presence of confinement in the relevant experimental setups, it is
crucial to be able to calculate these coefficients. This is a complicated task, because the explicit formula for the local exchange coefficients involves
computationally demanding $N-1$ dimensional integrals over an antisymmetrized $N$-body wave function (a Slater determinant) \cite{volosniev2014}.  In the present
paper, we describe a procedure for computing these coefficients that
bypasses the complexity of high-dimensional integrations, and
introduce associated software called CONAN that computes the
coefficients for up to 35 particles in arbitrary external potentials.

The source code to the CONAN program is freely available at 
\cite{codelink} including pre-compiled versions that run
out of the box.  
We ask that in any scientific publication based 
wholly or in part on CONAN, the
use of CONAN must be acknowledged and the present paper must be cited.

The paper is organized as follows. In Sec. \ref{sec:2} we introduce
the system and the spin chain Hamiltonian, $H_s$, that desribes it. In
Sec. \ref{sec:numerical-implementation} we propose an algorithm for
calculating parameters of $H_s$. In Sec. \ref{sec:units} we start the
presentation of CONAN by introducing the dimensionless variables that
our software uses. A guideline of how to use CONAN can be found in
Secs.~\ref{sec:guide} and \ref{sec:ex} and in the documentation
accompanying the program. Finally, in Sec. \ref{sec:concl} we give a
brief overview of the present study.

\section{The system}
\label{sec:2}
\begin{figure}[thbp]
  \centering
  
  \def\spinup at (#1,#2){\fill [blue!80!black] (#1,#2) circle (.13);
  \draw [->, thick] (#1,#2-.25) -- (#1,#2+.25);}
  \def\spindown at (#1,#2){\fill [red!80!black] (#1,#2) circle (.13);
  \draw [<-, thick] (#1,#2-.25) -- (#1,#2+.25);}
  \def\site at (#1,#2){\draw [densely dotted] (#1,#2) circle (.13);}

  \begin{tikzpicture}
    
    \begin{scope}[shift={(-.3,0)}] % particles in trap
      \shade [shading=radial,
      inner color=red!40!blue!70!white, outer color=white]
      (1.85,0) ellipse (1.7 and .3);

      \node [below] at (1.85,-.45) {$1/g \to 0$};
      
      \spinup at (.8,0);
      \spindown at (1.2,0);
      \spinup at (1.75,0);
      \spinup at (2.52,0);
      \spindown at (2.9,0);
    \end{scope}

    \begin{scope}[shift={(1.6,-1.3)}, xscale=1.4]; % trap potential
      \draw [black, thick, domain=-1.163:1.15, samples=100, smooth]
      plot (\x, {abs(\x)^6 + .3*cos((\x + 1) r)*cos((3*\x+.3) r)});
      \node [above] at (-1.163,2.3) {$V(x)$};
    \end{scope}

    \node at (4.2,0) {$\mapsto$};

    \begin{scope}[shift={(4.5,0)}] % particles on chain
      \spinup at (.8,0);
      \spindown at (1.2,0);
      \spinup at (1.75,0);
      \spinup at (2.52,0);
      \spindown at (2.9,0);
    \end{scope}

    \begin{scope}[shift={(4.5,-.5)}] % chain sites
      \site at (.8,0);
      \site at (1.2,0);
      \site at (1.75,0);
      \site at (2.52,0);
      \site at (2.9,0);

      \draw [thick] (.8+.13,0) -- (1.2-.13,0);
      \draw [thick] (1.2+.13,0) -- (1.75-.13,0);
      \draw [thick] (1.75+.13,0) -- (2.52-.13,0);
      \draw [thick] (2.52+.13,0) -- (2.9-.13,0);

      \node [below] at (1,-.05) {$\alpha_1$};
      \node [below] at (1.475,-.05) {$\alpha_2$};
      \node [below] at (2.135,-.05) {$\alpha_3$};
      \node [below] at (2.71,-.05) {$\alpha_4$};
    \end{scope}

  \end{tikzpicture}
  \caption{Illustration of the mapping for a system with the interaction strength, $g$. In the strong interaction
    regime, $1/g \to  0$, this system can be related (see the text) to a spin chain with trap-dependent exchange coefficients, $\alpha_k$.}
  \label{fig:map}
\end{figure}
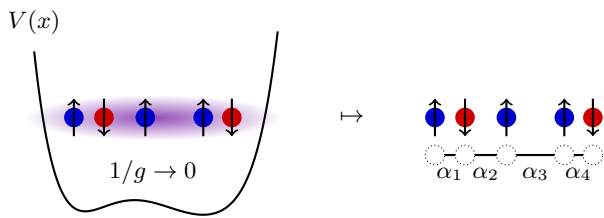

In this section we discuss the mapping of a strongly interacting
one-dimensional gas onto a spin chain, see
Fig.~\ref{fig:map}. A more detailed discussion of the mapping can be found in
Ref.~\cite{volosniev2015}.

\subsection{Confined strongly interacting gas}

We consider a system of particles in a one-dimensional trapping
potential $V(x)$. All particles are of the same mass, $M$, and divided
into two types which we call spin up and spin down. We denote by
$N_\uparrow$ ($N_\downarrow$) the number of spin up (spin down)
particles and by $N = N_\uparrow + N_\downarrow$ the total number of
particles. The spin up particles have coordinates $x_1, x_2, ...,
x_{N_\uparrow}$, whereas $x_{N_{\uparrow}+1}, ..., x_{N}$ are the
positions of the spin down particles. We postulate that a spin up
particle interacts with a spin down particle via a contact interaction
of strenth $g > 0$. Furthermore, we assume that other interactions
have strength $\kappa g$ with $\kappa > 0$. The dynamics of the system
is governed by the following Hamiltonian
\begin{align}
  \label{eq:hamiltonian}
  H = &\sum\limits_{i=1}^N H_0(x_i)
  + g \sum_{i=1}^{N_{\uparrow}}\sum_{j=N_{\uparrow}+1}^{N}
  \delta(x_i - x_j) \nonumber\\
  &+ \frac{\kappa g}{2} \sum_{i,j=1}^{N_{\uparrow}}
  \delta(x_i - x_j)
  + \frac{\kappa g}{2} \sum_{i,j=N_{\uparrow}+1}^N
  \delta(x_i - x_j)
  \; ,
\end{align}
where the single-particle Hamiltonian is given by
\begin{equation}
  \label{eq:single-particle-hamiltonian}
  H_0(x) = -\frac{1}{2} \frac{\partial^2}{\partial x^2} + V(x)
  \; ,
\end{equation}
where for simplicity we put $\hbar=M=1$. 
We assume that particles of the same type are bosons. However, from 
the limit $\kappa \rightarrow \infty$ one can also learn about fermionic systems, see, e.g., Ref.~\cite{volosniev2015}.

\subsection{Effective spin chain model}

Before we consider the case of strong interaction, i.e., $1/g \to 0$, we focus on the Tonks-Girardeau (TG) limit ($1/g = 0$). In the TG limit, the wavefunctions vanish whenever any two particles meet, $x_i = x_j$. Therefore, in this limit the particles cannot exchange their positions, and for each ordering of particles (e.g., $x_1<x_2<...<x_N$) the system is described with a wave function of spinless fermions $\Phi_0(x_1,\dots,x_N)$. For clarity from now on we assume that $\Phi_0$ corresponds to a ground state of spinless fermions. The discussion, however, can be extended easily to the excited manifolds. 
To construct $\Phi_0$ we find $\psi_i(x)$ -- the real eigenstates of $H_0$, i.e.,
\begin{equation}
  \label{eq:single-particle-se}
  H_0(x) \, \psi_i(x) = E_i \, \psi_i(x) \;, \quad i=1,2,\dots ,
\end{equation}
where $E_i$ are the corresponding eigenvalues.
From the $N$ lowest eigenstates, we construct the Slater determinant
wavefunction, $\Phi_0$. The corresponding energy
$E_0$ equals to the sum of the $N$ lowest single-particle energies, i.e., $E=\sum_{i=1}^N E_i$. 
Note that the $N$-body system has $N!/(N_\uparrow! \cdot N_\downarrow!)$ distinct orderings, and therefore the ground state manifold of $H$ is $N!/(N_\uparrow! \cdot N_\downarrow!)$-fold degenerate.

When the interaction strength is moved slightly away from the
TG limit, the wavefunctions become non-vanishing at $x_i = x_j$,
thus allowing the particles to hop. To linear order in
$1/g$, we can describe this hopping using a spin chain Hamiltonian with
nearest neighbor interaction:
\begin{equation}
  \label{eq:heisenberg}
  H_s = E_0 - \sum_{k=1}^{N-1} \frac{\alpha_k}{g} \left[ \frac{1}{2}
    \left(1 - \boldsymbol{\sigma}^k \cdot \boldsymbol{\sigma}^{k+1} \right)
    + \frac{1}{\kappa}
    \left(1 + \sigma_z^k \sigma_z^{k+1} \right) \right] \; ,    
\end{equation}
where $\boldsymbol{\sigma}^k = (\sigma_x^k, \sigma_y^k, \sigma_z^k)$
are the Pauli matrices acting on the spin of the particle at the $k$th
position and $\alpha_k$ are the exchange coefficients.  
In Ref.~\cite{volosniev2015}
it was shown that 
\begin{equation}
\alpha_k=\frac{\int_{x_1<x_2<\dots<x_{N-1}}\mathrm{d}x_1\dots\mathrm{d}x_{N-1} \left|\frac{\partial \Phi_0}{\partial x_N}\right|^2_{x_N=x_{k}}}{\int_{x_1<x_2<\dots<x_N}\mathrm{d}x_1\mathrm{d}x_2\dots\mathrm{d}x_N  |\Phi_0|^2}.
\label{eq:geometric-integral}
\end{equation}
For convenience we assume that $\Phi_0$ is normalized such that
$\int_{x_1<x_2<\dots<x_N}\mathrm{d}x_1\mathrm{d}x_2\dots\mathrm{d}x_N
|\Phi_0|^2=1$. The exchange coefficients $\alpha_k$ are called
geometric coefficients, derived from the fact that they only depend on
the geometry of the trap potential.

\section{Algorithm for calculating $\alpha_k$}
\label{sec:numerical-implementation}

Access to the coefficients $\alpha_k$ would effectively
solve the strongly interacting $N$-body trapped system, 
and, thus, we seek a way to compute them. In principle, one
could find the $N$ lowest energy single-particle wavefunctions
$\psi_i$, construct the Slater determinant wavefunction, $\Phi_0$, and
evaluate the $(N-1)$-dimensional integral in
Eq.~\eqref{eq:geometric-integral}.  However, multidimensional integrals 
are computationally demanding. Therefore, for more than just a few particles, we need
to cast $\alpha_k$ in a form better suited for numerical
calculations. In
Appendix~\ref{app:geometric-final} we show that $\alpha_k$ can be
expressed  as a sum of one-dimensional integrals
\begin{align}
  \label{eq:geometric-final}
  \alpha_k = \; &2
  \sum_{i=1}^N \sum_{j=1}^N \sum_{l=0}^{N-1-k}
  \frac{(-1)^{i+j+N-k}}{l!}
  {N-l-2 \choose k-1} \nonumber\\
  &\times \int_a^b \textrm{d}x \,
  \frac{2m}{\hbar^2} \big( V(x) - E_i \big) \, \psi_i (x) \,
  \frac{\textrm{d}\psi_j}{\textrm{d}x} \nonumber\\
  &\times
  \left[
    \frac{\partial^l}{\partial\lambda^l}
    \det \left[ (B(x) - \lambda \textbf{I})^{(ij)} \right]
  \right]_{\lambda = 0} \nonumber\\
  &+ \sum_{i=1}^N \left[ \frac{\textrm{d}\psi_i}{\textrm{d}x}
  \right]^2_{x=b}\;,
\end{align}
where \textbf{I} denotes the identity matrix, $(\;)^{(ij)}$ denotes
the $ij$'th submatrix obtained by removing the $i$'th
column and the $j$'th row, $B(x)$ is a $N \times N$ symmetric matrix
with the $mn$'th entry defined as the partial overlap of $\psi_m$ and
$\psi_n$, i.e.,
\begin{equation}
  \label{eq:B-matrix}
  [B(x)]_{m,n} = \int_a^x \textrm{d}y \, \psi_m(y) \, \psi_n(y) \; .
\end{equation}
The interval $[a,b]$ denotes the support of $V(x)$. For example, for a
harmonic oscillator potential the support is the entire $x$-axis,
i.e., $[a,b] = (-\infty,\infty)$; for a hard box, defined from $x=0$
to $x=L$, we have $[a,b] = [0,L]$.

In Eq.~\eqref{eq:geometric-final} the $(N-1)$-dimensional integral has
been rewritten as a sum of terms which scale more advantageously with
$N$. We note that similar reductions have been discussed in relation 
to calculating the densities of strongly interacting systems, see
Ref.~\cite{deuretz2008}.
Unfortunately, the new expression contains the $l$'th order
derivative in the square brackets which complicates numerical
calculations as $N$ increases. Therefore, we would like to simplify
the expression further.

\subsection{Simplifying the determinant}
From the standpoint of an effective numerical implementation of
Eq. \eqref{eq:geometric-final} the complicated part is the evaluation of the
derivatives of the determinant,
\begin{equation}
  \left[
    \frac{\partial^l}{\partial\lambda^l}
    \det\left[ (B(x) - \lambda\textbf{I})^{(ij)} \right]
  \right]_{\lambda=0} \; .
  \label{eq:derivative-of-det}
\end{equation}        
Our method for evaluating this expression is due to the fact that $ B
$ is symmetric (and real because we have chosen real wavefunctions
$\psi_i$), and, hence, diagonalizable using
an orthogonal matrix $ U = \left( \mathbf{u}_1 \ldots \mathbf{u}_N
\right) $ such that $ B = U^{\mathrm{T}} D U $, where $ D $ is a
diagonal matrix composed of the eigenvalues of $ B $.  We note that
taking the $ ij $'th submatrix of $B$ is equivalent to
removing a row and a column from $ U^{\mathrm{T}} $ and $ U $
respectively.  This observation allows us to show (see
Appendix~\ref{app:determinant} for details) that the expression in
Eq.~\eqref{eq:derivative-of-det} can be written as
\begin{equation}
  (-1)^{i+j} \, l! \,\mathbf{u}_j^{\mathrm{T}}
  \left(\sum_{n=0}^{l}p_{l-n} D^{-(n+1)}\right) \mathbf{u}_i \; ,
  \label{eq:dod-reexpressed}
\end{equation}
where $ p_k $ are the coefficients of the polynomial
\begin{equation}
  \det (D-\lambda\mathbf{I})
  = p_{N}\lambda^{N} + \ldots + p_1\lambda + p_0.
\end{equation}
Several comments are in order here. First of all, as $ D $ is
diagonal, it can be easily inverted as long as its entries are
nonzero. In Appendix~\ref{app:determinant} we prove that this is in
fact the case. Secondly, the coefficients $ p_k $ are easily
computable because $ \det (D-\lambda\mathbf{I}) $ is a determinant of
a diagonal matrix. Thirdly, a further reduction in computation
requirements can be achieved by doing the sum over $ l $ inside the
integral. Then we diagonalize $ B $ only once, rather than once for
each $ l $. Therefore, to evaluate the integrand we need to compute
the expression
\begin{align}
  \label{eq:det-final}
  & (-1)^{i+j}\sum\limits_{l=0}^{N-1-k} {N-2-l \choose k-1}
  \frac{1}{l!}  \left[ \frac{\partial^l}{\partial \lambda^l}
    \det\left[ (B - \lambda\textbf{I})^{(ij)} \right]
  \right]_{\lambda=0} \nonumber\\
  & = \mathbf{u}_j^{\mathrm{T}} \left[ \sum\limits_{l=0}^{N-1-k}
    {N-2-l \choose k-1} \sum\limits_{n=0}^l p_{l-n} D^{-(n+1)} \right]
  \mathbf{u}_i \; .
\end{align}
Finally, and perhaps the most interestingly, the expression in
Eq.~\eqref{eq:det-final} depends on $ i $ and $ j $ only through $
\mathbf{u}_i $ and $ \mathbf{u}_j $. Hence, the vast majority of the
computations are independent on $ i $ and $ j $ which prompts us to
take the sum over $ i $ and $ j $ inside the integral in
Eq.~\eqref{eq:geometric-final} as well, and reuse the result for the
derivatives of the determinant. The procedure reduces the computation time 
as the derivatives have to be computed only once for each $ x $ rather than computing it $ \sim N^2 $ times.

\subsection{Procedure}
Here we outline our method for computing the coefficients $\alpha_k$
from Eq. \eqref{eq:geometric-final}, where we first take all the sums
for a given $x$ and then perform the integration. To take the sums the
following is done:
\begin{itemize}
\item[1.] The entries of $ B(x) $ are evaluated. 
\item[2.] The matrix $ B(x) $ is diagonalized.
\item[3.] The coefficients $ p_k $ are computed.
\item[4.] The matrices $ D^{-(n+1)} $ are computed for all $ 0\leq
  n\leq N-1-k $ utilizing that $ D $ is diagonal.
\item[5.] The (diagonal) matrix inside the square brackets of
  Eq.~\eqref{eq:det-final} is evaluated.
\item[6.] The sum over $ i $ and $ j $ is taken. For this the product
  in Eq.~\eqref{eq:det-final} is multiplied by the appropriate factors
  of $ \psi_i $, $ \mathrm{d} \psi_j / \mathrm{d} x $, and $ (V-E_i)
  $.
\end{itemize}  
From the procedure sketched above we can estimate how the computation
time scales with the number of particles, $ N $. It seems that the
steps 2, 5, and 6 are the most demanding for large $N$, because the
computation time of all these steps scales roughly as $ O(N^3)
$. Still, this scaling is surprisingly good.

To further reduce the computation time we have exploited the fact that
modern computers have multiple computing cores. The easiest way to go
about parallelizing the program is to let each core compute a separate
geometric coefficient. In our implementation there is not a great
overlap in computation different coefficients, therefore, this way of
parallelizing leads to a much more effective use of the computing
power. Furthermore, the geometric coefficients only have a use when
all of them are known, so this parallelizing cannot slow down the
process.

\subsection{Arbitrary precision matrix computations}
There is one caveat associated with the outlined procedure which we
should like to bring attention to. The success of our algorithm relies
on an accurate diagonalization of $ B(x) $, which is composed of partial overlap integrals of the wave functions. Some
of these overlaps may be much smaller than the others depending on
which overlap and to what value of $ x $ it is being
evaluated. Therefore, the entries of $ B $ might span many orders of
magnitude, and the use of too small precision in the diagonalization
of $ B $ might yield a wrong outcome.

It turned out that the usual machine precision is insufficient for
$N\sim7$. This is likely due to an enhanced error because the
eigenvalues are taken to a high negative power in
Eq.~\eqref{eq:det-final}. We have found ourselves forced to implement
a diagonalization routine using numbers with until several thousands
 bits precision (for large $ N $) in order to determine the
eigenvalues and eigenvectors accurately.  The requirement for the
numerical precision to this diagonalization scales with the number of
particles so this inevitably influences the original estimate of an $
O(N^3) $ computation time. In Section~\ref{sec:ex} we will see that
the computation time typically scales with $O(N^{3.5\pm0.4})$.

\subsection{Computation of $B(x)$}
\label{sec:Box_exp}
The computation of $B(x)$ from Eq.~\eqref{eq:B-matrix} is
straightforward.  It requires only the eigenstates of the one-particle
Hamiltonian $H_0(x)$ from
Eq.~\eqref{eq:single-particle-hamiltonian}, and a
numerical procedure to perform the integration in
Eq.~\eqref{eq:B-matrix}. Both of these requirements can be fulfilled
by picking a suitable basis of states for the one-body problem. As
such a basis, we choose the eigenstates of a hard box potential on the
interval $[0,L]$, i.e.,
\begin{align}
  \label{eq:box-wavefunction}
  \phi_n(x) = \sqrt{\frac{2}{L}} \sin\left(\frac{n\pi x}{L}\right) \; ,
  \quad
  n  = 1,\dots, N_{b} \; ,
\end{align}
where $N_b$ defines the basis size used.
The basis of $N_b$ box wavefunctions in
Eq.~\eqref{eq:box-wavefunction} form an orthogonal basis on $[0,L]$,
which is complete in the limit $N_b \rightarrow \infty$. 

To solve the one-body problem accurately we need to pick the values
for $N_b$ and $L$.  The number of basis states should be
chosen much larger than the number of particles, $N_b \gg N$, such
that the higher momentum contributions are not important. For
instance, for the systems from Sec. \ref{sec:ex} correct results can
be obtained using a basis with around a few hundred elements.

The basis states are defined on a finite interval, therefore, $L$
should be chosen sufficiently large such that the wavefunctions,
$\psi_i$, are effectively confined to the region $[0,L]$. In other
words, putting $a=0$ and $b=L$ in Eq.~\eqref{eq:geometric-final}
should not change the value of $\alpha_k$ within a given precision.

After expressing the single-particle Hamiltonian, $H_0(x)$, as a $N_b \times
N_b$ matrix in the basis of box eigenstates, we find the expansion
coefficients for the wavefunctions, $\psi_i$, through the
usual diagonalization procedure. Let us denote with
\begin{equation}
  \label{eq:expansion-coefficient}
  [C]_{i,m} =  \int_0^L \textrm{d}x \,
  \phi_m(x) \psi_{i}(x),
\end{equation}
the expansion coefficients, and with $C$ the $N \times N_b$ matrix of
these. Then we derive the simple expression (see
Appendix~\ref{app:B_expanded}) for the matrix $B$
\begin{equation}
  \label{eq:B_expanded}
  B(x) = C f(x) C^\textrm{T} \; ,
\end{equation} 
where $f(x)$ is the matrix:
\begin{align}
  \label{eq:f-matrix}
  &\left[f(x)\right]_{m,n} =
\begin{cases}
  \frac{1}{\pi} \left[ \frac{1}{m-n}\sin
    \left((m-n)\frac{\pi x}{L}\right) \right. \\
  \hspace{5mm}
  - \left. \frac{1}{m+n}\sin \left((m+n)\frac{\pi x}{L}\right)\right]
  & \text{for } m\neq n \\
  \frac{1}{\pi} \left[ \frac{\pi x}{L} -
    \frac{1}{2m}\sin \left(\frac{2m \pi x}{L} \right) \right]
  & \text{for } m=n
\end{cases}
\end{align}
The elements of this matrix are less computationally expensive to
calculate than the original integrals in the $B(x)$-matrix, and the
coefficient matrix $C$ only needs to be calculated once. Calculating 
$B(x)$ by using Eq.~\eqref{eq:B_expanded} is therefore significantly 
faster than performing the $N^2$ integrals in Eq.~\eqref{eq:B-matrix} numerically. It should be noted that this 
increase in speed does not alter the way the computation time scales 
with the number of particles. Simply doing the $N^2$ integrals from 
Eq.~\eqref{eq:B-matrix} would ideally scale as 
$O(N^2)$, but in practice increases a bit faster due to the difficulty 
of integrating the rapidly oscillating excited states that come into 
play when $N$ becomes large. Similarly, calculating $B(x)$ using 
Eq.~\eqref{eq:B_expanded} requires 2 matrix products involving the 
$C$-matrix. This matrix has $N$ rows and $N_b$ columns, and so the 
calculation time scales as $O(N^2 N_b^2)$. Because the basis size does not need 
to be altered significantly as $N$ is increased, the scaling for this 
calculation is also close to $O(N^2)$.

\section{Units and change of units}
\label{sec:units}
Using the algorithm presented in the previous section we wrote a code
CONAN, which produces $\alpha_k$ for given $V(x)$ and $N$. We start
the presentation of our code by describing the conversion of the
relevant quantities into unitless numbers that CONAN works with.

When looking at the expressions above, only two types of dimensions appear: 
dimensions of length and dimensions of energy. Therefore, we need to specify 
what units are used for lengths and energies.  We let $\ell$ to be some unit of 
length, and define from this the unit of energy as
\begin{equation}
  \varepsilon = \frac{1}{2\ell^2} \; .
  \label{eq:varepsilon}
\end{equation}
Let the unitless variable that corresponds to a quantity $q$ be
denoted as $\tilde{q}$, i.e., $\tilde{x} = x/\ell$ and $\tilde{V}(x) =
V(x)/\varepsilon$. CONAN assumes that $\tilde x, \tilde V $ variables
are used in Eq. \eqref{eq:single-particle-hamiltonian}, and produces the corresponding
$\tilde{\alpha}_k$.  To convert $\tilde{\alpha}_k$ to $\alpha_k$ one
should use the equation
\begin{equation}
  \tilde{\alpha}_k = {\ell^3} \alpha_k \; ,
\label{eq:coefficient_units}
\end{equation}
which follows directly from our definition of dimensionless quantities. 

The procedure for translating from a physical system to the language of
the program and back works as follows:
\begin{itemize}
\item[1.] Pick a unit of length $\ell$.
\item[2.] Pick a sufficiently large box of length $\tilde{L}$.
\item[3.] Find the dimensionless potential as $\tilde V(\tilde x) =
  V(x/l)/\varepsilon$.
\item[4.] Enter $\tilde{L}$ and $\tilde{V}(\tilde{x})$ into the
  program and run it.
\item[5.] Convert $\tilde{\alpha}_k$ to $\alpha_k$ using Eq.~\eqref{eq:coefficient_units}.
\end{itemize}

As an example, let us consider a simple harmonic oscillator,
\begin{equation}
V(x)=\frac{\omega^2 x^2}{2} \; .
\end{equation}
For the concrete implementation in the program, we should shift it $x
\rightarrow x-L/2$ such that the minimum of $V(x)$ is in the center of
the box, but this matters not for the present discussion.

A reasonable choice for the unit of length and the corresponding
unit of energy is
\begin{align}
  \label{eq:ho-length-unit}
\ell = \sqrt{\frac{1}{\omega}} \; , \qquad
\varepsilon = \frac{1}{2} \omega \; .
\end{align}
With these definitions, the numerical potential to feed into the program
becomes
\begin{align}
\tilde{V}(\tilde{x}) & \equiv \frac{1}{\varepsilon} V(\tilde{x} \ell)= \tilde{x}^2 \; ,
\end{align}
and when the program finishes calculating the coefficients $\tilde{\alpha}_k$, 
these should be converted back to dimensionful quantities as
\begin{equation}
\alpha_k =  \omega ^{\frac{3}{2}} \tilde{\alpha}_k \; .
\end{equation}

With this example in mind, we can now approach the subject of how the
program behaves when we rescale the units. Assume we rescale the unit of
length by a factor $\delta$. Because the unit of energy is connected to
$\ell$ by Eq. \eqref{eq:varepsilon}, this would also rescale the unit
of energy. The scaling then is
\begin{align}
\ell \rightarrow \delta \ell \; , \qquad 
\varepsilon \rightarrow \delta^{-2} \varepsilon, \qquad
\tilde{V}(\tilde{x}) \rightarrow \delta^2 \tilde{V}(\delta \tilde{x})
\; .
\end{align}
Similarly, a rescaling of the energy by a factor $\lambda$ implies a
rescaling of the unit of length,
\begin{align}
  \varepsilon \rightarrow \lambda \varepsilon \; , \quad
  \ell \rightarrow \lambda^{-\frac{1}{2}} \ell \; , \quad
  \tilde{V}(\tilde{x})
  \rightarrow \lambda^{-1} \tilde{V}(\lambda^{-\frac{1}{2}}
  \tilde{x}) \; .
\end{align}
Note that these two types of scaling are identical up to a change of
notation if $\delta = \lambda^{-2}$. 

For the harmonic oscillator
in the example above, we see that changing the unit of energy
to $\gamma \omega/2$ yields the following change of the
unitless potential
\begin{align*}
  \tilde{V}(\tilde{x}) \rightarrow 
  \, \gamma^{-1} \tilde{V}(\gamma^{-\frac{1}{2}} \tilde{x})
  = \gamma^{-2} \tilde{x}^2 \; .
\end{align*}

The change of the units can be seen as the change of the size of the potential. For some potentials, this observation might be used to yield scaling laws for $\alpha_k$, as explored in the following section.

\subsection{Scaling of the coefficients with potential strength}
As mentioned above, the understanding of how the units enter in the
problem allows for the derivation of scaling laws for $\alpha_k$. To
see this, note first that besides the number of particles, the only
information about the system entered into the calculations is the
potential $\tilde{V}(\tilde{x})$. In other words, if we pick a
sufficiently large $\tilde{L}$ (and specify other
precision-parameters, as described in Section~\ref{sec:guide}), the
coefficients should depend only on the potential, i.e.,
\begin{equation}
  \tilde{\alpha}_k = \tilde{\alpha}_k [\tilde{V}(\tilde{x})] \; .
\end{equation}
Assume now that we are dealing with a potential which is a homogeneous
function, that is, a function for which there exists a point $a \in
(0,L)$ and a real number $s$, such that for any $k \in \mathbb{R}$ the
following holds:
\begin{equation}
  V(k(x-a)) = k^s V(x-a) \; .
\end{equation}
One example of such a potential is the harmonic oscillator ($s=2$)
introduced in the previous section. 

For potentials of this type, a scaling of the size of the potential
may be countered by a rescaling of $\ell$, so that the combinations of
the two scalings keep $\tilde{V}$ and, therefore, $\tilde{\alpha}$
unchanged. As shown in Appendix~\ref{sec:Appendix_Units_Scaling}, this
leads to the following scaling-law when the potential is scaled by a
factor $\gamma$,
\begin{equation}
\alpha_k [\gamma V] = \gamma^{\frac{3}{s+2}} \alpha_k [V] \; .
\label{eq:Scaling}
\end{equation}
For the harmonic case with $s=2$ the data from the program, run with
varying potential-sizes, confirms that the coefficients scale
as $\gamma^{\frac{3}{4}}$.

\section{Guide to the use of CONAN}
\label{sec:guide}
For a successful use of CONAN one has to understand in detail the input parameters given
to the program. In this section we list these parameters, and discuss
the means to estimate accuracy of results. To illustrate our code, in
Section~\ref{sec:ex} we consider two specific examples: a harmonic
potential and an asymmetric tilted potential in a box.

\begin{itemize}
\item[1.] The number of particles, $N$. Note that errors on some of
  the coefficients might become upredictable and large even for simple
  potentials, therefore, it is not advised to use $N > 35$. The option
  \texttt{-N} specifies the number of particles.
\item[2.] The smooth dimensionless trapping potential,
  $\tilde{V}(\tilde{x})$. This potential should trap the system on the
  interval $[0,\tilde L]$, i.e., the potential at the boundaries ($0$
  or $\tilde L$) of the box, that we use to expand the one-body wave
  functions, should be much larger than the typical energy scale. In
  other words the $N$ lowest-energy single-particle wavefunctions
  should be 'essentially' located in the interval $[0,\tilde L]$.  The
  option \texttt{-V} specifies the dimensionless potential as a
  mathematical function. The length of the box may be changed from the
  default value, $\tilde{L} = 100$, by submitting another value using
  the option \texttt{-L}.

  Note that a box trap should be chosen to be zero, i.e., $\tilde V(\tilde x)=0$. This    choise is required due to the numerical integration
  routines, that are used to calculate the matrix elements for the potential.
\item[3.] The number of basis states, $N_b$. This number should be
  chosen such that the $N$ lowest-energy single-particle wavefunctions
  are accurately represented in the basis of box states. The default
  value of the basis size in CONAN is $N_b = 300$. The option
  \texttt{-b} specifies this number to a different value.

  One can check whether $N_b$ is sufficiently large by comparing the
  results for different values of $N_b$, for instance $N_b = 200$,
  400, 600, 800; $N_b$ should be chosen in the range where the
  coefficients do not change with changing $N_b$ (within some desired
  precision).  Typically it is not a problem to choose $N_b$ larger
  than necessary. However, very large $N_b$ might lead to 
  instabilities in the numerical calculations of the potential in the
  basis of box states, $V_{nm} = \int_0^L \textrm{d}x \, \phi_n(x)
  \phi_m(x) V(x)$. These instabilities are due to quickly oscillating
  integrands in the matrix elements and can be avoided by adapting the
  intergration routines.

  Increasing the basis size increases the computation time through the
  diagonalization of the $N_b \times N_b$ Hamiltonian matrix, but this
  increase is almost independent on $N$. So if one is performing a few
  calculations for a large $N \sim 30$ system, where the computation
  time is an hour or so, it is a good idea to pick $N_b$ larger than
  necessary. On the other hand, if one is performing many calculations
  for small $N \sim 10$ with almost identical potentials (say, for
  noise studies as in Ref. \cite{loft2016}), then the computation time
  can be decreased by choosing the minimal $N_b$ that yields the
  desired accuracy.

\item[4.] The bit precision on the arbitrary precision calculations,
  $p$. The default value of $p$ is 256. If the bit precision is not
  sufficiently large, CONAN will run into an error or give
  coefficients which are several orders of magnitude different from
  each other. An accurate result is then obtained by increasing the
  value of $p$ using the option \texttt{-p}.  As a rule of thumb,
  always choose the bit precision as a power of 2, i.e., it should be
  increased in steps $p=$256, 512, 1024, 2048.  The computation time depends strongly on the bit
  precision (see the next section for an illustration), therefore, one
  does not want to set $p$ higher than necessary. 
  For just a few
  particles, $N \lesssim 5$, it suffices to use a bit precision
  smaller than the default value, say $p=64$ or $p=128$.
	
  While working with CONAN we noticed, that if a reasonable result can
  be obtained with the chosen bit precision, then typically this
  result is accurate. However, the accuracy can be confirmed only by increasing $p$.

\item[5.] The absolute or relative precision on the calculation of the
  single-particle wavefunctions and energies. These are changed using
  the options \texttt{--abs-solver} and \texttt{--rel-solver}
  respectively.  Other input parameters are the absolute or relative
  precision on the calculation of the integral in
  Eq.~\eqref{eq:geometric-final} (changed using \texttt{--abs-final}
  and \texttt{--rel-final}).  The default values of these parameters
  lead to accurate results for $N < 30$ for potentials similar to the
  presented in the next section.
\end{itemize}
The above list serves as a checklist one should review before
submitting an input to CONAN. There are other settings which can be
changed in order to obtain a specific output format or for parameter
dependent potentials, please consult the documentation for further
details.

\subsection{On accuracy of results}
\label{sec:accuracy}
Even if CONAN does not return an error, the resulting coefficients may
be wrong. Therefore, it is important to evaluate the result from the
program. First of all, one should ask oneself whether it makes sense
to interpret the coefficients as local exchange coefficients, provided
the trapping potential. Typically, the coefficients will resemble the
inverted potential $-V$, see for instance the examples in
Sec.~\ref{sec:ex} or Ref.~\cite{loft2016}. Next one should run CONAN again with an
increased basis size and/or precision, and compare the results. One
should keep increasing the basis size and/or precision until every
coefficient is determined to the desired degree of precision.

For $N>30$, we advise the user to be extra cautious because the
complexity of the calculations introduce errors in the results that
become harder and harder to avoid, even with an increased basis size
and precision parameters. Because the calculation time is also large,
it becomes very tedious to estimate the precision of the result by
peforming multiple calculations with different precision
settings. There is, however, a way to estimate the error on the
coefficients that works also for large values of $N$: We know that
reflecting the potential around its center point also
reflects the coefficients, i.e., $V(x) \mapsto V(L-x)$ yields $\alpha_k
\mapsto \alpha_{N-k}$. Thus comparing $\alpha_k$ calculated for $V(x)$
and $\alpha_{N-k}$ for $V(L-x)$ digit by digit, we can estimate the
error on the coefficients. This method works because the coefficients
we compare are calculated differently, i.e., we do not compare the
results of identical numerical routines. In the special case of a
symmetric potential, i.e., $V(x) = V(L-x)$, we can compare the
coefficients directly by checking to which degree of accuracy the
coefficients are symmetric, i.e., $\alpha_{N-k} = \alpha_k$.

We are hesitante to state something more specific about the precision of the
results, because every case should be considered separately. However, we do
note a few trends. For $N \leq 25$ high precision is only a matter of
picking $N_b$ and $p$ sufficiently large, then the error can be
reduced to less than 0.0001\% or 0.00001\%. Increasing
  the number of particles to $N \approx 30$ introduces a small error
  on a few coefficients of the order 0.001\% that cannot be easily reduced by
  changing $N_b$ or the precision parameters. We
  believe, these errors arise due to numerical instabilities, and we
  see no way to decrease them in the current version of the code. We
also note that the errors are not evenly distributed among the $N-1$
coefficients: a few coefficients gain larger errors while the
remaining coefficients seem accurate. This trend of non-uniformly
distributed errors continues as $N$ increases. For $N = 35$ the
largest errors are of the order 0.1\% to 1\%, depending on the
system. For $N$ around 40 the errors become unacceptably large, but
because the computation time is several hours for such large $N$, we
have not done any systematic studies in this range of $N$. Therefore,
we recommend to run CONAN for at most $N \approx 35$.
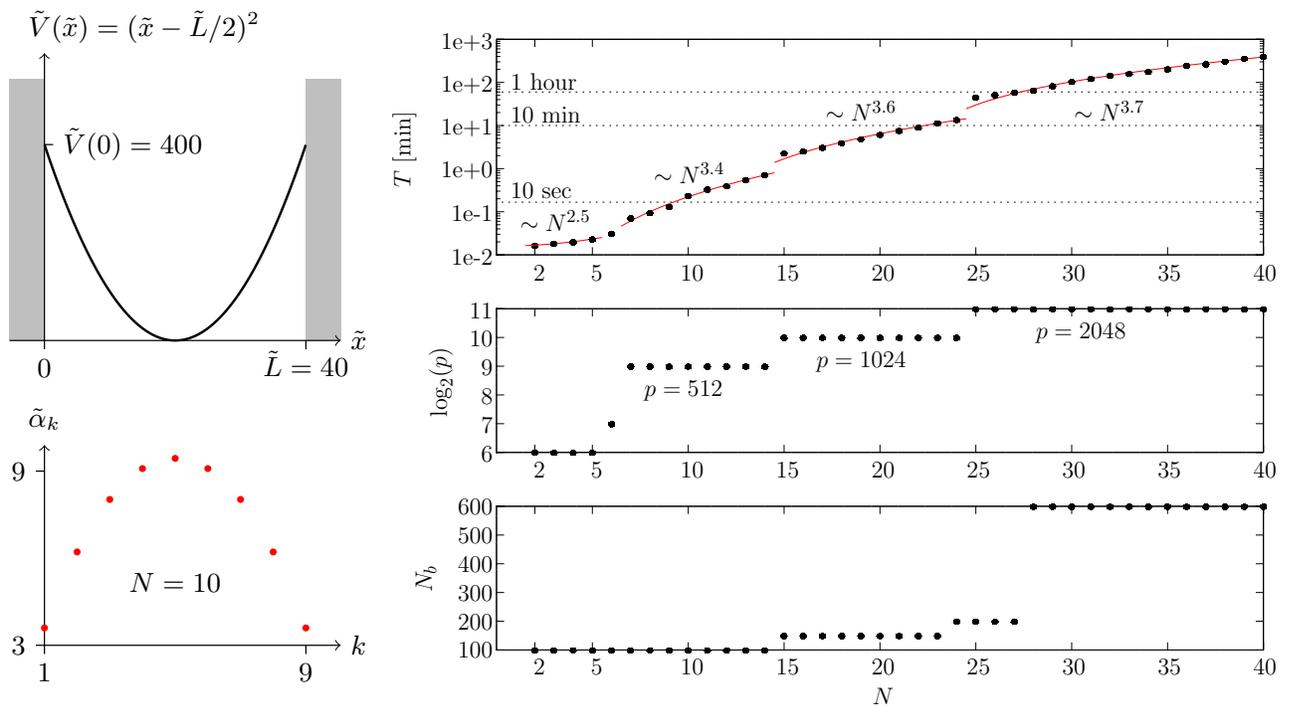
\begin{figure*}[tbp]
  \centering
  \resizebox{.5\columnwidth}{!}{
  \begin{picture}(100,100)(-10,-10)
    \begin{tikzpicture}
%      \draw [lightgray] (0,0) grid (4,4);
      \begin{scope}[shift={(.3,1.5)}];
      \fill [lightgray] (-.4,0) rectangle (0,3); % left wall
      \fill [lightgray] (3,0) rectangle (3.4,3); % right wall
      \draw [->] (-.4,0) -- (3.4,0); % 1. axis
      \node [right] at (3.4,0) {$\tilde x$};
      \node [below] at (0,-.1) {$0$};
      \draw (3,0) -- (3,-.1);
      \node [below] at (3,0) {$\tilde L = 40$};
      \draw [->] (0,-.1) -- (0,3.3); % 2. axis
      \node [right] at (-.3,3.6) {$\tilde V(\tilde x) = (\tilde x -
        \tilde L / 2)^2$};
      \draw [black, thick, domain=0:3, samples=100, smooth]
      plot (\x, {(\x - 1.5)^2});
      \draw (0,1.5^2) -- (.1,1.5^2);
      \node [right] at (.1,1.5^2) {$\tilde V(0) = 400$};
    \end{scope}
    % ./Conan -N 10 -V '(x-L/2)^2' -L 40 -p 512 -b 100
    % I = 3.59773 6.21586 8.02348 9.08789 9.43968 9.08789 8.02348 6.21586 3.59773
    \begin{scope}[shift={(.3,-2)}]
      \draw [->] (-.1,0) -- (3.4,0); % 1. axis
      \node [right] at (3.4,0) {$k$};
      \node [below] at (0,-.1) {$1$};
      \draw (3,0) -- (3,-.1);
      \node [below] at (3,-.1) {$9$};
      \draw [->] (0,-.1) -- (0,2.3); % 2. axis
      \node [right] at (-.3,2.6) {$\tilde \alpha_k$};
      \node [left] at (-.1,0) {$3$};
      \draw (-.1,2) -- (0,2);
      \node [left] at (-.1,2) {$9$};
      \node [above] at (1.5,.5) {$N = 10$};
      % data
      \fill [color=red] (0*3/8,3.59773*1/3-1) circle (.04);
      \fill [color=red] (1*3/8,6.21586*1/3-1) circle (.04);
      \fill [color=red] (2*3/8,8.02348*1/3-1) circle (.04);
      \fill [color=red] (3*3/8,9.08789*1/3-1) circle (.04);
      \fill [color=red] (4*3/8,9.43968*1/3-1) circle (.04);
      \fill [color=red] (5*3/8,9.08789*1/3-1) circle (.04);
      \fill [color=red] (6*3/8,8.02348*1/3-1) circle (.04);
      \fill [color=red] (7*3/8,6.21586*1/3-1) circle (.04);
      \fill [color=red] (8*3/8,3.59773*1/3-1) circle (.04);
    \end{scope}
    \end{tikzpicture}
  \end{picture}
  }
  \resizebox{1.5\columnwidth}{!}{
    \setlength{\unitlength}{1pt}
\begin{picture}(0,0)(-40,0)
\includegraphics{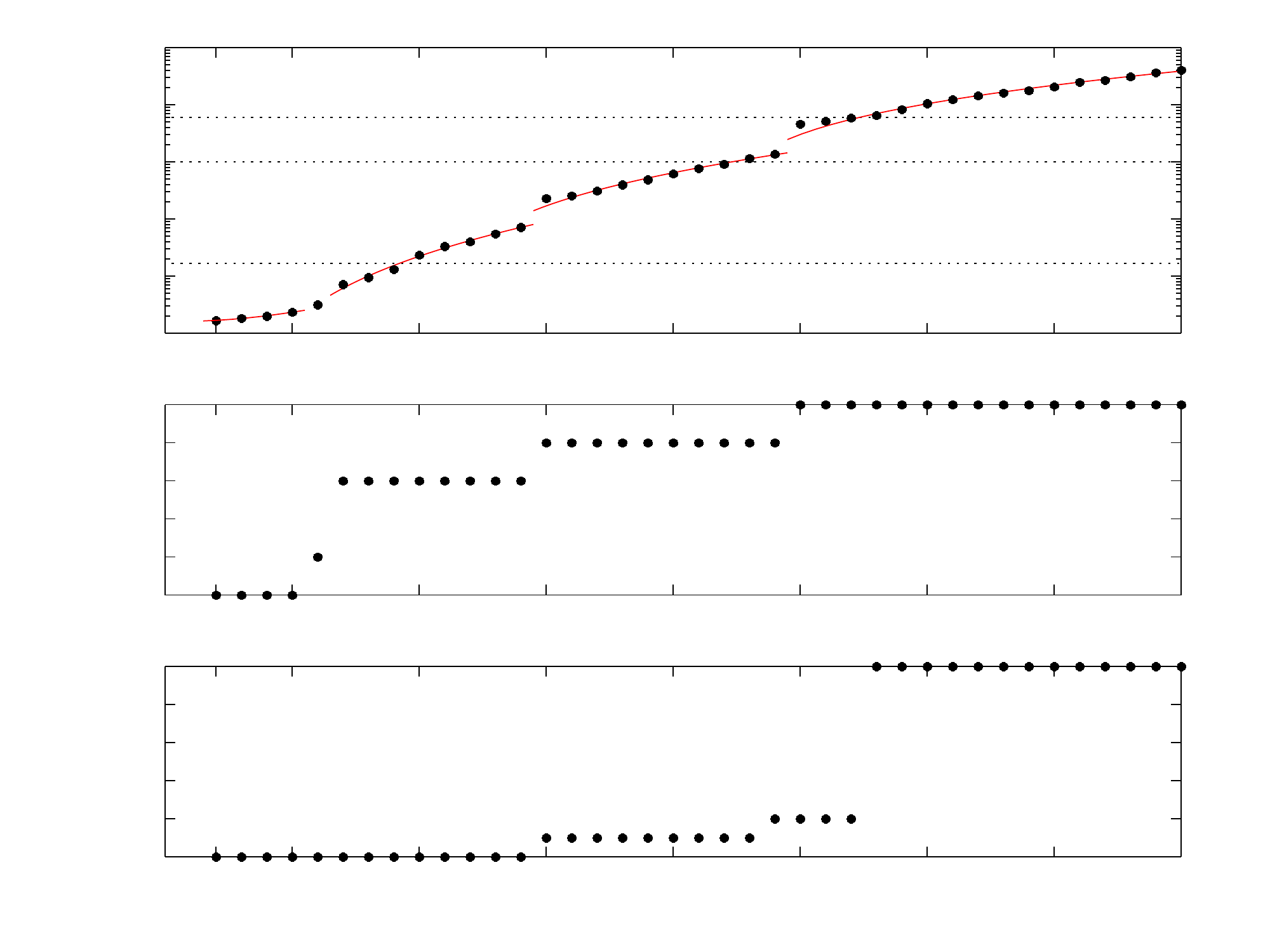}
\end{picture}
\begin{picture}(576,432)(-40,0)
  \fontsize{15}{0}
\selectfont\put(97.92,275.815){\makebox(0,0)[t]{\textcolor[rgb]{0,0,0}{{2}}}}
\fontsize{15}{0}
\selectfont\put(132.48,275.815){\makebox(0,0)[t]{\textcolor[rgb]{0,0,0}{{5}}}}
\fontsize{15}{0}
\selectfont\put(190.08,275.815){\makebox(0,0)[t]{\textcolor[rgb]{0,0,0}{{10}}}}
\fontsize{15}{0}
\selectfont\put(247.68,275.815){\makebox(0,0)[t]{\textcolor[rgb]{0,0,0}{{15}}}}
\fontsize{15}{0}
\selectfont\put(305.28,275.815){\makebox(0,0)[t]{\textcolor[rgb]{0,0,0}{{20}}}}
\fontsize{15}{0}
\selectfont\put(362.88,275.815){\makebox(0,0)[t]{\textcolor[rgb]{0,0,0}{{25}}}}
\fontsize{15}{0}
\selectfont\put(420.48,275.815){\makebox(0,0)[t]{\textcolor[rgb]{0,0,0}{{30}}}}
\fontsize{15}{0}
\selectfont\put(478.08,275.815){\makebox(0,0)[t]{\textcolor[rgb]{0,0,0}{{35}}}}
\fontsize{15}{0}
\selectfont\put(535.68,275.815){\makebox(0,0)[t]{\textcolor[rgb]{0,0,0}{{40}}}}
\fontsize{15}{0}
\selectfont\put(69.8822,280.8){\makebox(0,0)[r]{\textcolor[rgb]{0,0,0}{{1e-2}}}}
\fontsize{15}{0}
\selectfont\put(69.8822,306.72){\makebox(0,0)[r]{\textcolor[rgb]{0,0,0}{{1e-1}}}}
\fontsize{15}{0}
\selectfont\put(69.8822,332.64){\makebox(0,0)[r]{\textcolor[rgb]{0,0,0}{{1e+0}}}}
\fontsize{15}{0}
\selectfont\put(69.8822,358.56){\makebox(0,0)[r]{\textcolor[rgb]{0,0,0}{{1e+1}}}}
\fontsize{15}{0}
\selectfont\put(69.8822,384.48){\makebox(0,0)[r]{\textcolor[rgb]{0,0,0}{{1e+2}}}}
\fontsize{15}{0}
\selectfont\put(69.8822,410.4){\makebox(0,0)[r]{\textcolor[rgb]{0,0,0}{{1e+3}}}}
\fontsize{15}{0}
\selectfont\put(23.8822,345.6){\rotatebox{90}{\makebox(0,0)[b]{\textcolor[rgb]{0,0,0}{{$T$ [min]}}}}}
\fontsize{15}{0}
\selectfont\put(80.64,320.822){\makebox(0,0)[l]{\textcolor[rgb]{0,0,0}{{10 sec}}}}
\fontsize{15}{0}
\selectfont\put(80.64,366.363){\makebox(0,0)[l]{\textcolor[rgb]{0,0,0}{{10 min}}}}
\fontsize{15}{0}
\selectfont\put(80.64,386.532){\makebox(0,0)[l]{\textcolor[rgb]{0,0,0}{{1 hour}}}}
\fontsize{15}{0}
\selectfont\put(86.4,302.705){\makebox(0,0)[l]{\textcolor[rgb]{0,0,0}{{$\sim N^{2.5}$}}}}
\fontsize{15}{0}
\selectfont\put(167.04,331.454){\makebox(0,0)[l]{\textcolor[rgb]{0,0,0}{{$\sim N^{3.4}$}}}}
\fontsize{15}{0}
\selectfont\put(270.72,368.875){\makebox(0,0)[l]{\textcolor[rgb]{0,0,0}{{$\sim N^{3.6}$}}}}
\fontsize{15}{0}
\selectfont\put(420.48,368.875){\makebox(0,0)[l]{\textcolor[rgb]{0,0,0}{{$\sim N^{3.7}$}}}}
\fontsize{15}{0}
\selectfont\put(97.92,156.977){\makebox(0,0)[t]{\textcolor[rgb]{0,0,0}{{2}}}}
\fontsize{15}{0}
\selectfont\put(132.48,156.977){\makebox(0,0)[t]{\textcolor[rgb]{0,0,0}{{5}}}}
\fontsize{15}{0}
\selectfont\put(190.08,156.977){\makebox(0,0)[t]{\textcolor[rgb]{0,0,0}{{10}}}}
\fontsize{15}{0}
\selectfont\put(247.68,156.977){\makebox(0,0)[t]{\textcolor[rgb]{0,0,0}{{15}}}}
\fontsize{15}{0}
\selectfont\put(305.28,156.977){\makebox(0,0)[t]{\textcolor[rgb]{0,0,0}{{20}}}}
\fontsize{15}{0}
\selectfont\put(362.88,156.977){\makebox(0,0)[t]{\textcolor[rgb]{0,0,0}{{25}}}}
\fontsize{15}{0}
\selectfont\put(420.48,156.977){\makebox(0,0)[t]{\textcolor[rgb]{0,0,0}{{30}}}}
\fontsize{15}{0}
\selectfont\put(478.08,156.977){\makebox(0,0)[t]{\textcolor[rgb]{0,0,0}{{35}}}}
\fontsize{15}{0}
\selectfont\put(535.68,156.977){\makebox(0,0)[t]{\textcolor[rgb]{0,0,0}{{40}}}}
\fontsize{15}{0}
\selectfont\put(69.8822,162){\makebox(0,0)[r]{\textcolor[rgb]{0,0,0}{{6}}}}
\fontsize{15}{0}
\selectfont\put(69.8822,179.28){\makebox(0,0)[r]{\textcolor[rgb]{0,0,0}{{7}}}}
\fontsize{15}{0}
\selectfont\put(69.8822,196.56){\makebox(0,0)[r]{\textcolor[rgb]{0,0,0}{{8}}}}
\fontsize{15}{0}
\selectfont\put(69.8822,213.84){\makebox(0,0)[r]{\textcolor[rgb]{0,0,0}{{9}}}}
\fontsize{15}{0}
\selectfont\put(69.8822,231.12){\makebox(0,0)[r]{\textcolor[rgb]{0,0,0}{{10}}}}
\fontsize{15}{0}
\selectfont\put(69.8822,248.4){\makebox(0,0)[r]{\textcolor[rgb]{0,0,0}{{11}}}}
\fontsize{15}{0}
\selectfont\put(45.8822,205.2){\rotatebox{90}{\makebox(0,0)[b]{\textcolor[rgb]{0,0,0}{{$\log_2(p)$}}}}}
\fontsize{15}{0}
\selectfont\put(161.28,200.016){\makebox(0,0)[l]{\textcolor[rgb]{0,0,0}{{$p = 512$}}}}
\fontsize{15}{0}
\selectfont\put(264.96,217.296){\makebox(0,0)[l]{\textcolor[rgb]{0,0,0}{{$p = 1024$}}}}
\fontsize{15}{0}
\selectfont\put(397.44,234.576){\makebox(0,0)[l]{\textcolor[rgb]{0,0,0}{{$p = 2048$}}}}
\fontsize{15}{0}
\selectfont\put(97.92,38.1767){\makebox(0,0)[t]{\textcolor[rgb]{0,0,0}{{2}}}}
\fontsize{15}{0}
\selectfont\put(132.48,38.1767){\makebox(0,0)[t]{\textcolor[rgb]{0,0,0}{{5}}}}
\fontsize{15}{0}
\selectfont\put(190.08,38.1767){\makebox(0,0)[t]{\textcolor[rgb]{0,0,0}{{10}}}}
\fontsize{15}{0}
\selectfont\put(247.68,38.1767){\makebox(0,0)[t]{\textcolor[rgb]{0,0,0}{{15}}}}
\fontsize{15}{0}
\selectfont\put(305.28,38.1767){\makebox(0,0)[t]{\textcolor[rgb]{0,0,0}{{20}}}}
\fontsize{15}{0}
\selectfont\put(362.88,38.1767){\makebox(0,0)[t]{\textcolor[rgb]{0,0,0}{{25}}}}
\fontsize{15}{0}
\selectfont\put(420.48,38.1767){\makebox(0,0)[t]{\textcolor[rgb]{0,0,0}{{30}}}}
\fontsize{15}{0}
\selectfont\put(478.08,38.1767){\makebox(0,0)[t]{\textcolor[rgb]{0,0,0}{{35}}}}
\fontsize{15}{0}
\selectfont\put(535.68,38.1767){\makebox(0,0)[t]{\textcolor[rgb]{0,0,0}{{40}}}}
\fontsize{15}{0}
\selectfont\put(69.8822,43.2){\makebox(0,0)[r]{\textcolor[rgb]{0,0,0}{{100}}}}
\fontsize{15}{0}
\selectfont\put(69.8822,60.48){\makebox(0,0)[r]{\textcolor[rgb]{0,0,0}{{200}}}}
\fontsize{15}{0}
\selectfont\put(69.8822,77.76){\makebox(0,0)[r]{\textcolor[rgb]{0,0,0}{{300}}}}
\fontsize{15}{0}
\selectfont\put(69.8822,95.04){\makebox(0,0)[r]{\textcolor[rgb]{0,0,0}{{400}}}}
\fontsize{15}{0}
\selectfont\put(69.8822,112.32){\makebox(0,0)[r]{\textcolor[rgb]{0,0,0}{{500}}}}
\fontsize{15}{0}
\selectfont\put(69.8822,129.6){\makebox(0,0)[r]{\textcolor[rgb]{0,0,0}{{600}}}}
\fontsize{15}{0}
\selectfont\put(305.28,20.1767){\makebox(0,0)[t]{\textcolor[rgb]{0,0,0}{{$N$}}}}
\fontsize{15}{0}
\selectfont\put(35.8822,86.4){\rotatebox{90}{\makebox(0,0)[b]{\textcolor[rgb]{0,0,0}{{$N_b$}}}}}
\end{picture}
}
\caption{\textbf{Harmonic oscillator.} To the left we sketch
  the harmonic potential submitted to CONAN and the calculated
  geometric coefficients for $N=10$. To the right, on the top panel,
  we plot the computation time $T$ versus the number of particles $N$
  including the scaling for a fixed $p$. Below we show
  the chosen bit precision $p$ (middle) and basis size $N_b$
  (bottom). The errors on the calculated coefficients are below
  0.0001\% for $N < 29$ which increases to $\sim 0.1\%$ for $N=35$ and
  becomes even larger for larger $N$, see the main text for further
  discussion.}
  \label{fig:harmonic-ex}
\end{figure*}

\section{Examples}
\label{sec:ex}
Let us turn to two specific examples that illustrate the program.  In
these examples we calculate the geometric coefficients for various
values of $N$.  As $N$ increases we gradually increase the basis size
$N_b$ and bit precision $p$ in an effort to obtain high precision
results while keeping the computation time low. We run CONAN on a
computer with an Intel Xenon processor (CPU E5-2630 v3 @ 2.40 GHz
$\times$ 8) and note the computation time, $T$. To get an estimate of
how $T$ scales with $N$, we fit our data to the model $T = a + b \cdot
N^c$, where $a,b,c$ are fit parameters. We find that the computation
time typically scales as $O(N^{3.5 \pm 0.4})$ (depending on the
potential and range of $N$ values). We also find that a calculation
for $N=10$ particles takes approximately 10 seconds, and computations
for $N=20$ can be done in less than 10 minutes.

\begin{figure*}[tbp]
  \centering
  \resizebox{.5\columnwidth}{!}{
  \begin{picture}(100,100)(10,-10)
    \begin{tikzpicture}
%      \draw [lightgray] (0,0) grid (4,4);
      \begin{scope}[shift={(.3,1.5)}];
      \fill [lightgray] (-.4,0) rectangle (0,3); % left wall
      \fill [lightgray] (3,0) rectangle (3.4,3); % right wall
      \draw [->] (-.4,0) -- (3.4,0); % 1. axis
      \node [right] at (3.4,0) {$\tilde x$};
      \node [below] at (0,-.1) {$0$};
      \draw (3,0) -- (3,-.1);
      \node [below] at (3,0) {$\tilde L = 100$};
      \draw [->] (0,-.1) -- (0,3.3); % 2. axis
      \node [right] at (-.3,3.6) {$\tilde V(\tilde x) = 0.02 \,
        e^{-\tilde x  / 50}$};
      \draw [black, thick, domain=0:3, samples=100, smooth]
      plot (\x, {exp(-\x/1.5))});
      \draw (0,1) -- (.1,1);
      \node [right] at (.05,1.05) {$0.02$};
      \draw [<->] (1.1,.135335) -- (1.1,1);
      \node [right] at (1.1,.6) {$\approx 0.017$};
    \end{scope}
    % ./Conan -N 10 -V '0.02*exp(-x/50)' -p 512 -b 200
    % I 0.00344217 0.00358365 0.00370033 0.0037967 0.00387638 0.00394235 0.00399703 0.00404238 0.00408004
    \begin{scope}[shift={(.3,-2)}]
      \draw [->] (-.1,0) -- (3.4,0); % 1. axis
      \node [right] at (3.4,0) {$k$};
      \node [below] at (0,-.1) {$1$};
      \draw (3,0) -- (3,-.1);
      \node [below] at (3,-.1) {$9$};
      \draw [->] (0,-.1) -- (0,2.3); % 2. axis
      \node [right] at (-.3,2.6) {$\tilde \alpha_k$};
      \node [left] at (-.1,0) {$.0034$};
      \draw (-.1,2) -- (0,2);
      \node [left] at (-.1,2) {$.0041$};
      \node [above] at (1.5,.3) {$N = 10$};
      % data
      \fill [color=red] (0*3/8,0.00344217*2/0.0007-2*0.0034/0.0007) circle (.04);
      \fill [color=red] (1*3/8,0.00358365*2/0.0007-2*0.0034/0.0007) circle (.04);
      \fill [color=red] (2*3/8,0.00370033*2/0.0007-2*0.0034/0.0007) circle (.04);
      \fill [color=red] (3*3/8,0.0037967*2/0.0007-2*0.0034/0.0007) circle (.04);
      \fill [color=red] (4*3/8,0.00387638*2/0.0007-2*0.0034/0.0007) circle (.04);
      \fill [color=red] (5*3/8,0.00394235*2/0.0007-2*0.0034/0.0007) circle (.04);
      \fill [color=red] (6*3/8,0.00399703*2/0.0007-2*0.0034/0.0007) circle (.04);
      \fill [color=red] (7*3/8,0.00404238*2/0.0007-2*0.0034/0.0007) circle (.04);
      \fill [color=red] (8*3/8,0.00408004*2/0.0007-2*0.0034/0.0007) circle (.04);
    \end{scope}
    \end{tikzpicture}
  \end{picture}
  }
  \resizebox{1.5\columnwidth}{!}{
    \setlength{\unitlength}{1pt}
\begin{picture}(0,0)(-40,0)
\includegraphics{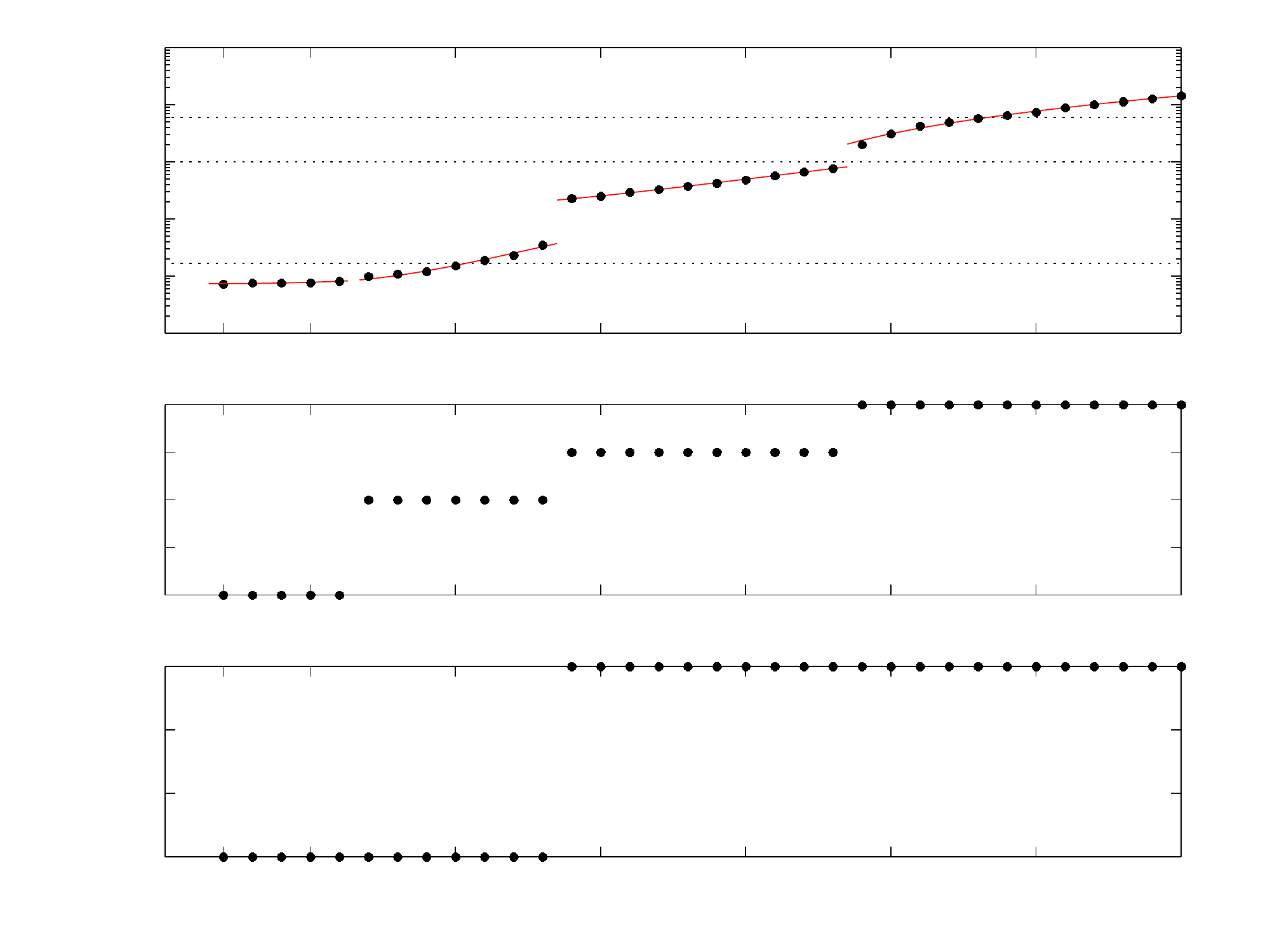}
\end{picture}%
\begin{picture}(576,432)(-40,0)
\fontsize{15}{0}
\selectfont\put(101.211,275.815){\makebox(0,0)[t]{\textcolor[rgb]{0,0,0}{{2}}}}
\fontsize{15}{0}
\selectfont\put(140.709,275.815){\makebox(0,0)[t]{\textcolor[rgb]{0,0,0}{{5}}}}
\fontsize{15}{0}
\selectfont\put(206.537,275.815){\makebox(0,0)[t]{\textcolor[rgb]{0,0,0}{{10}}}}
\fontsize{15}{0}
\selectfont\put(272.366,275.815){\makebox(0,0)[t]{\textcolor[rgb]{0,0,0}{{15}}}}
\fontsize{15}{0}
\selectfont\put(338.194,275.815){\makebox(0,0)[t]{\textcolor[rgb]{0,0,0}{{20}}}}
\fontsize{15}{0}
\selectfont\put(404.023,275.815){\makebox(0,0)[t]{\textcolor[rgb]{0,0,0}{{25}}}}
\fontsize{15}{0}
\selectfont\put(469.851,275.815){\makebox(0,0)[t]{\textcolor[rgb]{0,0,0}{{30}}}}
\fontsize{15}{0}
\selectfont\put(535.68,275.815){\makebox(0,0)[t]{\textcolor[rgb]{0,0,0}{{35}}}}
\fontsize{15}{0}
\selectfont\put(69.8822,280.8){\makebox(0,0)[r]{\textcolor[rgb]{0,0,0}{{1e$-$2}}}}
\fontsize{15}{0}
\selectfont\put(69.8822,306.72){\makebox(0,0)[r]{\textcolor[rgb]{0,0,0}{{1e$-$1}}}}
\fontsize{15}{0}
\selectfont\put(69.8822,332.64){\makebox(0,0)[r]{\textcolor[rgb]{0,0,0}{{1e+0}}}}
\fontsize{15}{0}
\selectfont\put(69.8822,358.56){\makebox(0,0)[r]{\textcolor[rgb]{0,0,0}{{1e+1}}}}
\fontsize{15}{0}
\selectfont\put(69.8822,384.48){\makebox(0,0)[r]{\textcolor[rgb]{0,0,0}{{1e+2}}}}
\fontsize{15}{0}
\selectfont\put(69.8822,410.4){\makebox(0,0)[r]{\textcolor[rgb]{0,0,0}{{1e+3}}}}
\fontsize{15}{0}
\selectfont\put(23.8822,345.6){\rotatebox{90}{\makebox(0,0)[b]{\textcolor[rgb]{0,0,0}{{$T$ [min]}}}}}
\fontsize{15}{0}
\selectfont\put(81.4629,320.822){\makebox(0,0)[l]{\textcolor[rgb]{0,0,0}{{10 sec}}}}
\fontsize{15}{0}
\selectfont\put(81.4629,366.363){\makebox(0,0)[l]{\textcolor[rgb]{0,0,0}{{10 min}}}}
\fontsize{15}{0}
\selectfont\put(81.4629,386.532){\makebox(0,0)[l]{\textcolor[rgb]{0,0,0}{{1 hour}}}}
\fontsize{15}{0}
\selectfont\put(105,293.167){\makebox(0,0)[l]{\textcolor[rgb]{0,0,0}{{$\sim N^{3.0}$}}}}
\fontsize{15}{0}
\selectfont\put(180.04,328.625){\makebox(0,0)[l]{\textcolor[rgb]{0,0,0}{{$\sim N^{4.3}$}}}}
\fontsize{15}{0}
\selectfont\put(290,368.875){\makebox(0,0)[l]{\textcolor[rgb]{0,0,0}{{$\sim N^{3.9}$}}}}
\fontsize{15}{0}
\selectfont\put(440,368.875){\makebox(0,0)[l]{\textcolor[rgb]{0,0,0}{{$\sim N^{3.1}$}}}}
\fontsize{15}{0}
\selectfont\put(101.211,156.977){\makebox(0,0)[t]{\textcolor[rgb]{0,0,0}{{2}}}}
\fontsize{15}{0}
\selectfont\put(140.709,156.977){\makebox(0,0)[t]{\textcolor[rgb]{0,0,0}{{5}}}}
\fontsize{15}{0}
\selectfont\put(206.537,156.977){\makebox(0,0)[t]{\textcolor[rgb]{0,0,0}{{10}}}}
\fontsize{15}{0}
\selectfont\put(272.366,156.977){\makebox(0,0)[t]{\textcolor[rgb]{0,0,0}{{15}}}}
\fontsize{15}{0}
\selectfont\put(338.194,156.977){\makebox(0,0)[t]{\textcolor[rgb]{0,0,0}{{20}}}}
\fontsize{15}{0}
\selectfont\put(404.023,156.977){\makebox(0,0)[t]{\textcolor[rgb]{0,0,0}{{25}}}}
\fontsize{15}{0}
\selectfont\put(469.851,156.977){\makebox(0,0)[t]{\textcolor[rgb]{0,0,0}{{30}}}}
\fontsize{15}{0}
\selectfont\put(535.68,156.977){\makebox(0,0)[t]{\textcolor[rgb]{0,0,0}{{35}}}}
\fontsize{15}{0}
\selectfont\put(69.8822,162){\makebox(0,0)[r]{\textcolor[rgb]{0,0,0}{{7}}}}
\fontsize{15}{0}
\selectfont\put(69.8822,183.6){\makebox(0,0)[r]{\textcolor[rgb]{0,0,0}{{8}}}}
\fontsize{15}{0}
\selectfont\put(69.8822,205.2){\makebox(0,0)[r]{\textcolor[rgb]{0,0,0}{{9}}}}
\fontsize{15}{0}
\selectfont\put(69.8822,226.8){\makebox(0,0)[r]{\textcolor[rgb]{0,0,0}{{10}}}}
\fontsize{15}{0}
\selectfont\put(69.8822,248.4){\makebox(0,0)[r]{\textcolor[rgb]{0,0,0}{{11}}}}
\fontsize{15}{0}
\selectfont\put(45.8822,205.2){\rotatebox{90}{\makebox(0,0)[b]{\textcolor[rgb]{0,0,0}{{$\log_2(p)$}}}}}
\fontsize{15}{0}
\selectfont\put(110,174.96){\makebox(0,0)[l]{\textcolor[rgb]{0,0,0}{{$p = 64$}}}}
\fontsize{15}{0}
\selectfont\put(185,187.92){\makebox(0,0)[l]{\textcolor[rgb]{0,0,0}{{$p = 512$}}}}
\fontsize{15}{0}
\selectfont\put(290,209.52){\makebox(0,0)[l]{\textcolor[rgb]{0,0,0}{{$p = 1024$}}}}
\fontsize{15}{0}
\selectfont\put(420,231.12){\makebox(0,0)[l]{\textcolor[rgb]{0,0,0}{{$p = 2048$}}}}
\fontsize{15}{0}
\selectfont\put(101.211,38.1767){\makebox(0,0)[t]{\textcolor[rgb]{0,0,0}{{2}}}}
\fontsize{15}{0}
\selectfont\put(140.709,38.1767){\makebox(0,0)[t]{\textcolor[rgb]{0,0,0}{{5}}}}
\fontsize{15}{0}
\selectfont\put(206.537,38.1767){\makebox(0,0)[t]{\textcolor[rgb]{0,0,0}{{10}}}}
\fontsize{15}{0}
\selectfont\put(272.366,38.1767){\makebox(0,0)[t]{\textcolor[rgb]{0,0,0}{{15}}}}
\fontsize{15}{0}
\selectfont\put(338.194,38.1767){\makebox(0,0)[t]{\textcolor[rgb]{0,0,0}{{20}}}}
\fontsize{15}{0}
\selectfont\put(404.023,38.1767){\makebox(0,0)[t]{\textcolor[rgb]{0,0,0}{{25}}}}
\fontsize{15}{0}
\selectfont\put(469.851,38.1767){\makebox(0,0)[t]{\textcolor[rgb]{0,0,0}{{30}}}}
\fontsize{15}{0}
\selectfont\put(535.68,38.1767){\makebox(0,0)[t]{\textcolor[rgb]{0,0,0}{{35}}}}
\fontsize{15}{0}
\selectfont\put(69.8822,43.2){\makebox(0,0)[r]{\textcolor[rgb]{0,0,0}{{200}}}}
\fontsize{15}{0}
\selectfont\put(69.8822,72){\makebox(0,0)[r]{\textcolor[rgb]{0,0,0}{{300}}}}
\fontsize{15}{0}
\selectfont\put(69.8822,100.8){\makebox(0,0)[r]{\textcolor[rgb]{0,0,0}{{400}}}}
\fontsize{15}{0}
\selectfont\put(69.8822,129.6){\makebox(0,0)[r]{\textcolor[rgb]{0,0,0}{{500}}}}
\fontsize{15}{0}
\selectfont\put(305.28,20.1767){\makebox(0,0)[t]{\textcolor[rgb]{0,0,0}{{$N$}}}}
\fontsize{15}{0}
\selectfont\put(35.8822,86.4){\rotatebox{90}{\makebox(0,0)[b]{\textcolor[rgb]{0,0,0}{{$N_b$}}}}}
\end{picture}
}
\caption{\textbf{Asymmetric tilted potential in a box.} To the left we sketch the potential submitted to CONAN and the calculated
  geometric coefficients for $N=10$. To the right, on
  the top panel, we plot the computation time $T$ versus the number of
  particles $N$ including the scaling for a fixed $p$. Below we show the chosen bit precision $p$ (middle) and
  basis size $N_b$ (bottom). The estimated error is 0.0001\% or less for $N
  \lesssim 28$, but increases to at most 3\% for $N=35$.}
  \label{fig:ex-tilt}
\end{figure*}

\subsection{Harmonic potential}
\label{sec:ex-harmonic}
Let us first calculate the geometric coefficients for a harmonic
oscillator potential
\begin{equation}
  \label{eq:ex-harmonic}
  V(x) = \frac{1}{2} \omega^2 x^2 \;,
\end{equation}
using the typical oscillator units given in
Eqs.~\eqref{eq:ho-length-unit}. Note that these coefficients
 for $N \leq 30$ were already presented in Ref.~\cite{loft2015}. Using the
oscillator units introduced in Section~\ref{sec:units}, we write the
dimensionless potential
\begin{equation}
  \tilde V(\tilde x) = \tilde x^2 \; .
\end{equation}
However, we cannot submit the above potential to CONAN because the
potential has its minimum at the boundary of the box, so we shift the
minimum of the potential to the center of the box, 
\begin{equation}
  \label{eq:ex-harmonic-dimless}
  \tilde V(\tilde x) = (\tilde x - \tilde L / 2)^2 \; .
\end{equation}
At the boundaries of the box, the above potential yields $\tilde V (0)
= \tilde V( \tilde L) = \tilde L^2 / 4$, so picking $\tilde L = 40$
leads to the value of the potential at the boundary being 400.  This
number is much larger than any other energy scale in the problem,
which is what we want.

We calculate the geometric coefficients for $N = 2,\dots,40$ with
gradually increasing basis size $N_b$ and bit precision $p$. An input
line to the program may look like\\
\texttt{ ./Conan -V '(x-L/2)\^{}2' -L 40 -N 8 -p 512 -b 100}

The basis size, bit precision and computation time for $N=2,\dots,40$
are shown in Figure~\ref{fig:harmonic-ex}. In the top panel of the
figure we show the computation time $T$ versus $N$. We see that the
computation time increases noticably when $p$ is
increased. For every plateaux characterized with $p = 64$, 512,
1024 and 2048, we calculate how the computation time scales with $N$
and see that it typically scales with $O(N^{3.5 \pm 0.2})$.

The accuracy of these results can be estimated from
the calculations with a larger number of basis states and higher
precision parameters up to around 30 particles. Because the potential is
symmetric around the middle of the box, we may also estimate the
precision by checking to which degree the coefficients are symmetric,
$\alpha_k \stackrel{?}{=} \alpha_{N-k}$.

For $N < 29$ the estimated in this way error on the coefficients is less than 0.0001\%. For $N=30$ the error on a few coefficients increases to 0.001\%, while most of the coefficients are symmetric to a higher degree of precision. For $N=33$ the error on the most imprecise
coefficients is $\sim 0.01\%$, growing to $\sim 0.1\%$ for $N=35$. It
is worth noticing that the precision on the $N=35$ result is not
increased by increasing the basis size to $N_b = 800$ and the absolute
and relative integral precision to 1e-08 using the options
\texttt{--abs-final} and \texttt{--rel-final}, indicating
  that the precision cannot become better for these large $N$ results
 in the current version of CONAN due to numerical
  instabilities as discussed in the final paragraph of
  Section~\ref{sec:accuracy}. Pushing the limits of CONAN's
capabilities by increasing $N$ further, we note that the error
increases to a few percent for $N=37$ and up to the worst case
deviations $\sim 100\%$ for $N=40$. Clearly, one cannot blindly trust
results for these large values of $N$, and we believe that CONAN
should be run for no more than approximately 35 particles.

\subsection{Asymmetric tilted potential in a box}
\label{sec:ex-tilt}
Here we consider a simple box potential with an added exponentially
decaying potential. Therefore, the resulting potential is not symmetric
around the center of the box, but tilted to one side, i.e.,
\begin{equation}
  \label{eq:ex-tilt}
   V (x) =
   \begin{cases}
     0.02 \varepsilon\, e^{-\frac{x}{ 50 \ell}} & \mbox{if } 0 < x < L=100 \ell \\
     \infty & \mbox{otherwise}
   \end{cases}
\end{equation}
where the numerical factors have been chosen such that the difference
$V(0) - V( L) \approx 0.017 \varepsilon$ is comparable with the energy
scale of the system, and the energy unit $\varepsilon$ used by CONAN
is given by Eq.~\eqref{eq:varepsilon}. 

We submit to CONAN the
following dimensionless potential
\begin{equation}
  \label{eq:tilt-dimensionless}
  \tilde V (\tilde x ) = 0.02 \, e^{-\tilde x / 50} \;.
\end{equation}
We calculate the geometric coefficients for $N=2,\dots, 35$ with the
basis size $N_b$ and bit precision $p$ chosen to retain a high degree
of precision on the results. The corresponding computation time is shown in Fig.~\ref{fig:ex-tilt} (top) for $p = 64$, 512, 1024 and 2048.

We found that $N_b = 200$ was sufficient to ensure an error below
0.0001\% for $N < 14$,  for more particles the basis size had to been increased
 to retain a high degree of precision. By comparing
$\alpha_k$ with $\alpha_{N-k}$ calculated for the mirror reflected
potential $V(L-x)$, we estimate an error of 0.0001\% for $N=28$ for
the most inaccurate coefficients. This increases to 0.001\% for $N=30$
and 3\% for $N=35$.

\section{Conclusions}
\label{sec:concl}
We presented an algorithm for computing the local exchange
coefficients for a strongly interacting $N$-particle system confined
to one dimension by an external potential. We discussed the numerical
implementation of the method in a piece of open source software,
CONAN. Then, we discussed the use of the program, including examples
and estimates of the precision of the results. We found that CONAN
could produce reliable results for up to around $N = 35$, and that the
computation time typically scales as $O(N^{3.5 \pm 0.4})$. Computation
times were around 10 seconds for $N=10$ and less than 10 minutes for
$N=20$. The approach described here may be extended in a straightforward
manner to compute, for instance, densities in strongly interacting
systems or correlation functions.

\paragraph*{Remark:} 
While preparing this paper we became aware of a recent paper by
Deuretzbacher {\it et al.} \cite{deuretzbacher2016} which presents a
method for computing the coefficients that has some similarities to
the one used here. We note that while Deuretzbacher {\it et al.}
use a fitting method with Chebyshev polynomials to get high-order
derivatives, we use a recursive formula as described in the appendix.
This could have influence on the numerical stability of either
algorithm. It would be very interesting to explore whether one may
combine these two different approach to achieve even larger stability
than currently available in either approach.

\section*{Acknowledgments}
We would like to thank M. Valiente, A.~S.~Jensen, D.~V.~Fedorov,
O.~V.~Marchukov, and D.~Petrosyan for collaboration on the spin model
approach.  We thank E.~M.~Eriksen and E.~J.~Lindgren for discussions
on early developments on a formula to compute the local exchange
coefficients, as well as J.~M.~Midtgaard and A.~A.~S.~Kalaee for help
in testing these developments in spin systems. A.~G.~V.  and
N.~T.~Z. acknowledge support from the Danish Council for Independent
Research DFF Natural Sciences Sapere Aude program, and
A.~G.~V. acknowledges partial support from the Helmholtz Association
under contract HA216/EMMI. N.~J.~S.~L., L.~B.~K., A.~E.~T., and
N.~T.~Z. acknowledge support by grants from the Carlsberg
Foundation. The development of CONAN was assisted by J.~Termansen and
J.~H.~Jensen with support from the Carlsberg Foundation.

\newpage
\onecolumngrid
\appendix

%\section{Derivation of Eq.~\eqref{eq:geometric-final}}
\section{Derivation of Eq.~(6)}
\label{app:geometric-final}
In this appendix we derive Eq.~\eqref{eq:geometric-final} from
Eq.~\eqref{eq:geometric-integral}. We start by writing the Slater
determinant that defines the wave function for the system of $N$
spinless fermions using the Leibniz formula for determinants
\begin{equation}
\Phi_0(x_1,\dots,x_N)=\sum_{\pi}\mathrm{sign}(\pi)\prod_{i=1}^N\psi_{\pi(i)}(x_i),
\end{equation}
here $\pi$ denotes the permutation operator that acts on the set of
first $N$ natural numbers $\{1,2,\dots,N\}$, and $\psi_j$ is the $j$th
normalised one-body wave function defined in the main text.  Next, we
write $\alpha_k$ as
\begin{equation}
\alpha_k=\sum\limits_{i=1}^N \sum\limits_{j=1}^N(-1)^{i+j}\int_{a}^{b}\mathrm{d}x_k
\frac{\partial \psi_i(x_k)}{\partial x_k}
\frac{\partial \psi_j(x_k)}{\partial x_k}\int_{a\leq x_1<x_2<\dots<x_{N-1} \leq b} 
\mathrm{d}x_1\dots\mathrm{d}x_{k-1}\mathrm{d}x_{k+1}\dots\mathrm{d}x_{N-1}
(\xi^i\xi^j)(x_1,\dots,x_{N-1}),
\end{equation}
where the interval $[a,b]$ is the support of the trapping potential, see the main text for details. 
To obtain this expression we used the Laplace expression for $\Phi_0$,
\begin{equation}
\Phi_0=\sum_{i=1}^N (-1)^{i+N} \psi_i(x_N) \xi^i(x_1,\dots,x_{N-1}),
\end{equation}
where $\xi^i(x_1,\dots,x_{N-1}) =
\sum_{\pi_i}\mathrm{sign}(\pi_i)\prod_{j=1}^{N-1}
\psi_{\pi_i(j)}(x_j)$, $\pi_i$ is the permutation operator defined on
the set of $N-1$ elements: $\{1,\dots,N\} \setminus i$.  To produce
Eq.~\eqref{eq:geometric-final} we simplify the inner integral
\begin{equation}
I_{k,ij}=\int_{x_1<x_2<\dots<x_{N-1}} 
\mathrm{d}x_1\dots\mathrm{d}x_{k-1}\mathrm{d}x_{k+1}\dots\mathrm{d}x_{N-1}
(\xi^i\xi^j)(x_1,\dots,x_{N-1}).
\end{equation}
For this we note that $\xi^i\xi^j$ is a symmetric function, i.e.,
$(\xi^i\xi^j)(\dots,x_k,\dots,x_l,\dots)=(\xi^i\xi^j)(\dots,x_l,\dots,x_k,\dots)$,
which allows us to change the integration limits as
\begin{align}
I_{k,ij}=\frac{1}{(k-1)!(N-1-k)!}
\int_{a}^{x_k} \mathrm{d}x_1 \int_{a}^{x_k} \mathrm{d}x_2 \dots 
\int_{a}^{x_k} \mathrm{d}x_{k-1} \int_{x_k}^{b} \mathrm{d}x_{k+1}  \dots \int_{x_k}^{b} 
\mathrm{d}x_{N-1}\xi^i\xi^j.
\label{inner1}
\end{align}

Throughout our investigation we noticed that similar integrals appear
also for observables. Therefore, we find it useful to consider a more general integral, that has a similar structure
\begin{equation}
P^n_l(c)[F\Phi] = \int_{a}^c \mathrm{d}x_1
\dots\int_{a}^c \mathrm{d}x_{l-1}\int_{c}^b
\mathrm{d}x_{l}\dots\int_{c}^b \mathrm{d}x_{n} F(x_1,\dots,x_n)\Phi(x_1,\dots,x_n) \; ,
\end{equation}
where $l-1$ integrals should be taken from $a$ to $c$, other integrals
are taken from $c$ to $b$, the superscript $n$ defines the number of
variables. Functions $F$ and $\Phi$ are determinants build with the
orthonormalized sets of functions $\{f_i\}$ and $\{\phi_i\}$
correspondingly.
That is

$
\Phi({x}_1, {x}_2, \ldots, {x}_n) =
\left|
   \begin{matrix} \phi_1({x}_1) & \phi_2({x}_1) & \cdots & \phi_n({x}_1) \\
                      \phi_1({x}_2) & \phi_2({x}_2) & \cdots & \phi_n({x}_2) \\
                      \vdots & \vdots & \ddots & \vdots \\
                      \phi_1({x}_n) & \phi_2({x}_n) & \cdots & \phi_n({x}_n)
   \end{matrix} \right|,
$ and 
$
F({x}_1, {x}_2, \ldots, {x}_n) =
\left|
   \begin{matrix} f_1({x}_1) & f_2({x}_1) & \cdots & f_n({x}_1) \\
                      f_1({x}_2) & f_2({x}_2) & \cdots & f_n({x}_2) \\
                      \vdots & \vdots & \ddots & \vdots \\
                      f_1({x}_n) & f_2({x}_n) & \cdots & f_n({x}_n)
   \end{matrix} \right|.
$

\noindent Let us first consider $P^n_{n+1}(c)[F\Phi]$. Noticing that both $F$ and $\Phi$ are fully antisymmetric functions in their
variables, we write $P^n_{n+1}$ as
\begin{align}
P^n_{n+1}(c)[F\Phi] = n! \int_{a}^c \mathrm{d}x_{1}\dots\int_{a}^c
\mathrm{d}x_{n} f_1(x_1) f_2(x_2)\dots f_n(x_n)\left|
   \begin{matrix} \phi_1({x}_1) & \phi_2({x}_1) & \cdots & \phi_n({x}_1) \\
                      \phi_1({x}_2) & \phi_2({x}_2) & \cdots & \phi_n({x}_2) \\
                      \vdots & \vdots & \ddots & \vdots \\
                      \phi_1({x}_n) & \phi_2({x}_n) & \cdots & \phi_n({x}_n)
   \end{matrix} \right|  \nonumber \\
= n! \int_{a}^c \mathrm{d}x_{1}\dots\int_{a}^c \mathrm{d}x_{n}  \left|
   \begin{matrix} f_1(x_1) \phi_1({x}_1) & f_1(x_1) \phi_2({x}_1) & \cdots & f_1(x_1) \phi_n({x}_1) \\
                      f_2(x_2) \phi_1({x}_2) & f_2(x_2) \phi_2({x}_2) & \cdots & f_2(x_2) \phi_n({x}_2) \\
                      \vdots & \vdots & \ddots & \vdots \\
                      f_n(x_n)\phi_1({x}_n) & f_n(x_n)\phi_2({x}_n) & \cdots & f_n(x_n)\phi_n({x}_n)
   \end{matrix} \right|\;,
\end{align}
the integration can be performed inside of the determinant, which
leads to $ P^n_{n+1}(c)[F\Phi]= n! \mathrm{det} A(c)$ where the matrix
elements in $A(c)$ are defined as $[A(c)]_{i,j} = \int_a^c \mathrm{d}x f_i(x)
\phi_j(x)$.

Next we consider $P^n_n$
\begin{align}
P^n_n(c)[F\Phi] &= \int_{a}^c \mathrm{d}x_1 \dots
\int_{a}^c \mathrm{d}x_{n-1}\int_{c}^b
\mathrm{d}x_{n} F(x_1,\dots,x_n)\Phi(x_1,\dots,x_n) \nonumber \\
&=
\sum\limits_{i=1}^n\sum\limits_{j=1}^{n}
(-1)^{i+j}\int_c^{b}\mathrm{d}x_{n}f_i(x_n)\phi_j(x_n)
P^{n-1}_n(c)[F^i\Phi^j] \nonumber\\
&= (n-1)! \sum\limits_{i=1}^n\sum\limits_{j=1}^{n} (-1)^{i+j}\int_c^{b}
\mathrm{d}x_{n}f_i(x_n)\phi_j(x_n) \mathrm{det} \left[(A(c))^{(ij)} \right],
\label{eq:appendix_a8}
\end{align}
where the functions $F^i$ and $\Phi^j$ are obtained from the Laplace
expansions of $F$ and $\Phi$ respectively, the matrix $(A(c))^{(ij)}$
is obtained from the matrix $A(c)$ by crossing out the $i$'th row and
$j$'th column (i.e., $A^{(ij)}$ is the $ij$'th submatrix). From here
we can proceed by two different paths: First we can notice that the
expression in Eq. \eqref{eq:appendix_a8} can be rewritten using
Jacobi's formula for the derivative of the determinant as $\mathrm{tr}
(T {\rm adj}(A)) = \frac{\partial \mathrm{det}(A+\lambda T)}{\partial
  \lambda}\big|_{\lambda=0}$, where the matrix $T$ is defined as
$[T]_{i,j}=\int_c^b \mathrm{d}x f_i(x)\phi_j(x)$, and ${\rm adj} A$
denotes the adjugate of $A$. This observation allows us to derive the
identity $P^n_n(c)[F\Phi]=(n-1)!\frac{\partial \mathrm{det}(A+\lambda
  T)}{\partial \lambda}\big|_{\lambda=0}$. By repeating the same steps
(i.e., first is to write $P^n_j$ using $P^{n-1}_{j}$, second is to use
Jacobi's formula) for $P^n_{n-1},P^n_{n-2},\dots$ we obtain
$P^n_l(c)[F\Phi]=(l-1)!\frac{\partial^{n-l+1} \mathrm{det}( A+\lambda
  T)}{\partial \lambda^{n-l+1}}\big|_{\lambda=0}$. 

Another path, which was used to derive Eq.~\eqref{eq:geometric-final},
rests on the assumption that $f_i$ and $\phi_j$ are orthonogonal to
each other, i.e., $\int_a^b \mathrm{d}x f_i(x)\phi_j(x)=\delta_{ij}$,
where $\delta_{ij}$ is Kronecker's delta.  This assumption allows us
to rewrite Eq.~\eqref{eq:appendix_a8} as
\begin{align}
P^n_n(c)[F\Phi]
&= (n-1)!\sum\limits_{i=1}^n\sum\limits_{j=1}^{n}(-1)^{i+j} \mathrm{det} A^{(ij)}
(\delta_{ij}-[A]_{i,j}) \nonumber\\
&=-(n-1)!\left(\frac{\partial \mathrm{det}(A-\lambda
    \textbf{I})}{\partial \lambda}\bigg|_{\lambda=0}+n \mathrm{det}A\right).
\label{eq:appendix_a9}
\end{align}
Let us now consider $P^n_{n-1}(c)$
\begin{align}
P^n_{n-1}(c)[F\Phi]
&=
\sum\limits_{i=1}^n\sum\limits_{j=1}^{n}
(-1)^{i+j}\int_c^{b}\mathrm{d}x_{n}f_i(x_n)\phi_j(x_n)
P^{n-1}_{n-1}(c)[F^i\Phi^j] \nonumber\\
&= \sum\limits_{i=1}^n\sum\limits_{j=1}^{n}
P^{n-1}_{n-1}(c)[F^i\Phi^j] \delta_{ij} - P^n_n(c)[F\Phi]\nonumber \\
&= (n-2)!\left[\frac{\partial^2
    \mathrm{det}(A-\lambda \textbf{I})}{\partial \lambda^2}
  \bigg|_{\lambda=0}+(n-1)\frac{\partial
    \mathrm{det}(A-\lambda \textbf{I})}{\partial
    \lambda}\bigg|_{\lambda=0}\right]
+(n-1)!\left[\frac{\partial \mathrm{det}
    (A-\lambda \textbf{I})}{\partial \lambda}\bigg|_{\lambda=0}
  +n \mathrm{det}A\right].
\end{align} 
The pattern for $P^n_{l}(c)$ can be guessed now, and we write $P^n_l$ as
\begin{equation}
P^n_l(c)[F\Phi]=(-1)^{n+1-l}
\sum_{i=0}^{n+1-l} (n-i)!  {n+1-l \choose i}
\frac{\partial^i \mathrm{det}(A(c)-\lambda \textbf{I})}{\partial \lambda^i}\bigg|_{\lambda=0}.
\label{eq:Pnlfinal}
\end{equation}
One can easily check, e.g., by induction, that this expression is
correct.  The meaning of the factors in the sum can be understood from
the expression for $P^n_{n-1}$; the $(n-i)!$ factor always comes with
the $i$th derivative, whereas the ${n+1-l \choose i}$ factor
shows how many terms with the $i$'th derivative are in
the expression.

Now, we use the expression for $P_l$ to obtain  $\alpha_k$. 
Firstly, we write $I_{k,ij}$ as 
\begin{align}
I_{k,ij}=\frac{1}{(k-1)!(N-1-k)!}
\int_{a}^{x_k} \mathrm{d}x_1 \int_{a}^{x_k} \mathrm{d}x_2 \dots 
\int_{a}^{x_k} \mathrm{d}x_{k-1} \int_{x_k}^{b} \mathrm{d}x_{k+1}  \dots \int_{x_k}^{b} 
\mathrm{d}x_{N-1}F(x_1,...,x_{N-1})\Phi(x_1,...,x_{N-1}),
\end{align}
where the functions $F$ and $\Phi$ are defined above, assuming that
$n=N-1$, and
$\{f_1,...,f_{N-1}\}=\{\psi_1,...,\psi_{i-1},\psi_{i+1},...,\psi_N\}$
and
$\{\phi_1,...,\phi_{N-1}\}=\{\psi_1,...,\psi_{j-1},\psi_{j+1},...,\psi_N\}$. Secondly,
we rewrite $I_{k,ij}$ as
\begin{align}
I_{k,ij}=\sum_{l,m} (-1)^{l+m}\frac{f_l(x_k)\phi_m(x_k)}{(k-1)!(N-1-k)!} 
\int_{a}^{x_k} \mathrm{d}x_1 \int_{a}^{x_k} \mathrm{d}x_2 \dots 
\int_{a}^{x_k} \mathrm{d}x_{k-1} \int_{x_k}^{b} \mathrm{d}x_{k+1}  \dots \int_{x_k}^{b} 
\mathrm{d}x_{N-1}F^l\Phi^m,
\end{align}
where $\Phi^m$ $(F^m)$ is obtained from $\Phi$ $(F)$ by crossing out the $k$th row and $m$th column.  The integral can be easily taken, using the results from Eq. \ref{eq:Pnlfinal}, i.e., 
\begin{align}
I_{k,ij}=\sum_{l,m} \frac{f_l(x_k)\phi_m(x_k)}{(k-1)!(N-1-k)!} 
(-1)^{N-k-1+l+m} \sum_{r=0}^{N-k-1} (N-2-r)!  {N-k-1 \choose r}
\frac{\partial^r \mathrm{det}((A(x_k))^{(lm)}-\lambda
  \textbf{I}^{(ij)})}{\partial \lambda^r}\bigg|_{\lambda=0},
\end{align}
where we use $\textbf{I}^{(ij)}$, because in this case $\int_a^b \mathrm{d}x f_l(x)\phi_m(x)=[\textbf{I}^{(ij)}]_{l,m}$. Now we note that $\phi_m(x)f_l(x)=\frac{\partial [A(x)]_{l,m}}{\partial x}$ and use Jacobi's formula to obtain the expression for $\alpha_k$
\begin{align}
  \label{eq:geometric-pre-bt}
  \alpha_k =
  \sum_{i=1}^N \sum_{j=1}^N \sum_{l=0}^{N-1-k}
  \frac{(-1)^{i+j+N-1-k}}{l!}
  {N-l-2 \choose k-1}
  \int_a^b \textrm{d}x \,
  \frac{\textrm{d}\psi_i}{\textrm{d}x}
  \frac{\textrm{d}\psi_j}{\textrm{d}x}
  \frac{\textrm{d}}{\textrm{d}x}
  \left[
    \frac{\partial^l}{\partial\lambda^l}
    \det \left[ (B(x) - \lambda \textbf{I})^{(ij)} \right]
  \right]_{\lambda = 0} \; ,
\end{align}
where $B(x)$ is the symmetric matrix defined in Eq.~\eqref{eq:B-matrix}.
Next we integrate by parts to eliminate the derivative of the
expression in square brackets. This procedure yields
\begin{align}
  \label{eq:geometric-with-bt}
  \alpha_k = \; &2
  \sum_{i=1}^N \sum_{j=1}^N \sum_{l=0}^{N-1-k}
  \frac{(-1)^{i+j+N-k}}{l!}
  {N-l-2 \choose k-1}
  \int_a^b \textrm{d}x \,
  \frac{2m}{\hbar^2} \big( V(x) - E_i \big) \, \psi_i (x) \,
  \frac{\textrm{d}\psi_j}{\textrm{d}x}
  \left[
    \frac{\partial^l}{\partial\lambda^l}
    \det \left[ (B(x) - \lambda \textbf{I})^{(ij)} \right]
  \right]_{\lambda = 0} + \mathcal{B}\; ,
  \end{align}
  where $\mathcal{B}$ denotes the boundary term arising from the
  partial integration
 \begin{align}
   \label{eq:bt-start}
  \mathcal{B} =
  \sum\limits_{i=1}^N \sum\limits_{j=1}^N \sum\limits_{l=0}^{N-1-k}
  \frac{(-1)^{i+j+N-1-k}}{l!} {N-l-2 \choose k-1}
  \left.
    \frac{\textrm{d}\psi_i}{\textrm{d}x}
    \frac{\textrm{d}\psi_j}{\textrm{d}x}
    \left[
      \frac{\partial^l}{\partial\lambda^l}
      \det \left[ (B(x) - \lambda \textbf{I})^{(ij)} \right]
    \right]_{\lambda = 0}
  \right|_a^b \; .
\end{align}
Note that to obtain this equation we use the Schr\"odinger equation
from Eq.~\eqref{eq:single-particle-se}. The boundary term in
Eq. \eqref{eq:bt-start} can be simplified significantly. To do this we
consider
\begin{equation}
  \label{eq:deriv-det}
  K_ {ij}(x) =
      \left[
      \frac{\partial^l}{\partial\lambda^l}
      \det \left[ (B(x) - \lambda \textbf{I})^{(ij)} \right]
    \right]_{\lambda = 0} \;,
\end{equation}
evaluated at the boundary points $x=a$ and $x=b$, where the matrix
$B(x)$ reduces to $B(a) = \textbf{0}$ or $B(b) =
\textbf{I}$. Therefore, $K_{ij}(x)$ at $x=a$ and $x=b$ can be written
as
\begin{equation}
  K_{ij}(a)=\left[
      \frac{\partial^l}{\partial\lambda^l}
\det \left[ ( - \lambda \textbf{I})^{(ij)} \right] \right]_{\lambda = 0}, \qquad  K_{ij}(b)=\left[
      \frac{\partial^l}{\partial\lambda^l}
\det \left[ ((1 - \lambda) \textbf{I})^{(ij)} \right] \right]_{\lambda = 0}.
\end{equation}
If $i=j$ then $\textbf{I}^{(ij)}$ is the $(N-1) \times (N-1)$ identity
matrix, however if $i \neq j$ the matrix has a zero-row (and
zero-column), so its determinant is $0$. Thus,
$\mathrm{det}\textbf{I}^{(ij)}$ is a Kronecker delta,
$\delta_{ij}$. This observation leads to the following expressions
\begin{equation}
\label{eq:deriv-det_1}
 K_{ij}(a)=\left[
      \frac{\partial^l}{\partial\lambda^l}
( - \lambda)^{N-1} \delta_{ij}\right]_{\lambda = 0}, \;  K_{ij}(b)=\left[
      \frac{\partial^l}{\partial\lambda^l}
(1 - \lambda)^{N-1} \delta_{ij}\right]_{\lambda = 0}\;.
\end{equation}
Now let us evaluate these expressions. At the lower limit we have
\begin{align}
  \label{eq:deriv-det-a}
  K_{ij}(a) =
   \left.
    (-1)^l \frac{(N-1)!}{(N-l-1)!} ( - \lambda)^{N-l-1} \delta_{ij}
  \right|_{\lambda=0}
  = (-1)^l \frac{(N-1)!}{(N-l-1)!} \delta_{N-l-1,0} \, \delta_{ij} \; .
\end{align}
Notice that $N-l-1 \geq N-(N-1-k)-1 \geq N- (N-2) - 1 = 1 > 0$ for all
 terms in the sum over $l$ in Eq.~\eqref{eq:bt-start}, so
$K_{ij}(a) = 0$ there. At the upper limit we
have
\begin{align*}
  K_{ij}(b) =
  \left.
    (-1)^l \frac{(N-1)!}{(N-l-1)!} (1 - \lambda)^{N-l-1} \delta_{ij}
  \right|_{\lambda=0}
  = (-1)^l \frac{(N-1)!}{(N-l-1)!} \delta_{ij} \; .
\end{align*}
Inserting this result into Eq.~\eqref{eq:bt-start}, we obtain the
boundary term in the form
\begin{align*}
  \mathcal{B} = \sum\limits_{i=1}^N
  \left[ \frac{\textrm{d}\psi_i}{\textrm{d}x} \right]_{x=b}^2
  \sum\limits_{l=0}^{N-1-k} (-1)^{N-1-k+l}
  {N-l-2 \choose k-1} \frac{(N-1)!}{(N-l-1)! \, l!} \; ,
\end{align*}
which simplifies as the sum over $l$ equals unity. This can be proven
using the binomial theorem and the definition of the beta function.
\begin{align*}
  &\sum\limits_{l=0}^{N-1-k} (-1)^{N-1-k+l} {N-l-2 \choose k-1}
  \frac{(N-1)!}{(N-l-1)! \, l!}\\
  &= \frac{(N-1)!}{(N-1-k)!(k-1)!}
  \sum\limits_{l=0}^{N-1-k} (-1)^{N-1-k+l} {N-1-k \choose l}
  \frac{1}{N-l-1} \\
  &= \frac{(N-1)!}{(N-1-k)!(k-1)!}
  \sum\limits_{l=0}^{N-1-k} (-1)^{N-1-k+l} {N-1-k \choose l}
  \int_0^1 \mathrm{d}x \, x^{N-l-2}\\
  &= \frac{(N-1)!}{(N-1-k)!(k-1)!}
  \int_0^1 \mathrm{d}x \, x^{k-1}
  \sum\limits_{l=0}^{N-1-k} {N-1-k \choose l}
  (-x)^{N-1-k-l}\\
  &= \frac{(N-1)!}{(N-1-k)!(k-1)!}
  \int_0^1 \mathrm{d}x \, x^{k-1} (1-x)^{N-k-1}\\
  &= 1 \; ,
\end{align*}
where we have recognized the last integral as the beta function. Thus,
we finally have derived Eq.~\eqref{eq:geometric-final}.

%\section{Derivation of Eq.~\eqref{eq:det-final}}
\section{Derivation of Eq.~(11)}
\label{app:determinant}
\noindent Here we prove the equality   
\begin{equation}
\label{eq:det11}
  \left. \frac{\partial^l}{\partial\lambda^l}
    \det\left[ (B(x) - \lambda\textbf{I})^{(ij)} \right] \right|_{\lambda=0} =
  (-1)^{i+j} l!\mathbf{u}_j^{\mathrm{T}} \left(\sum_{n=0}^{l}p_{l-n}
    D^{-(n+1)}\right) \mathbf{u}_i,
\end{equation}
presented in Sec.~\ref{sec:numerical-implementation}.

By choosing real valued wave functions one can ensure that $ B $ is a
real and symmetric matrix. By the spectral theorem it is then possible
to diagonalize it using an orthogonal matrix $ U = \left( \mathbf{u}_1
  \ldots \mathbf{u}_N \right) $ such that $ B = U^{\mathrm{T}} D U $
with $ D $ being a diagonal matrix containing the eigenvalues of $ B $. Let us denote
by $ U_n = \left( \mathbf{u}_1 \ldots \mathbf{u}_{n-1} \;
  \mathbf{u}_{n+1} \ldots \mathbf{u}_N \right)$ , i.e., $U_n$ is the matrix $ U $ with
the $ n $'th column removed. In this notation the $ ij $'th submatrix
in Eq. \eqref{eq:det11} is simply
\begin{equation}
  (B-\lambda \mathbf{I})^{(ij)}
  = \left(U^{\mathrm{T}}(D-\lambda\mathbf{I})U\right)^{(ij)}
  = U_i^{\mathrm{T}} (D-\lambda\mathbf{I}) U_j \; .
\end{equation}
To proceed further we note that it is possible to turn $ U_i^{\mathrm{T}} $ and $ U_j $ into square matrices (by inserting respectively an extra row and column) without changing the value of the
determinant. This is possible when $ D-\lambda\mathbf{I} $ is
invertible in a small region around $ \lambda = 0 $ which is the case
as long as all eigenvalues of $ B $ are non-zero. In this case
\begin{equation}
  \begin{pmatrix}
    \mathbf{u}_i^{\mathrm{T}} \\
    U_i^{\mathrm{T}}
  \end{pmatrix}
  (D-\lambda\mathbf{I})
  \begin{pmatrix}
    [(D-\lambda\mathbf{I})^{-1}\mathbf{u}_i] & U_j
  \end{pmatrix}
  =
  \begin{pmatrix}
    1 & \mathbf{u}_i^{\mathrm{T}}(D-\lambda\mathbf{I}) U_j \\
    \mathbf{0} & (B-\lambda \mathbf{I})^{(ij)}
  \end{pmatrix} \; ,
  \label{eq:squareMatrices}
\end{equation}
where we used that $ U $ is an orthogonal matrix. One can prove that
the determinant is indeed unchanged by expanding the determinant in
the first column of the RHS of the above expression.  As the product
matrices are now square the determinant can be evaluated using the
product rule for determinants, hence
\begin{equation}
  \det \left[ (B-\lambda \mathbf{I})^{(ij)} \right]
  = \det
  \begin{pmatrix}
    \mathbf{u}_i^{\mathrm{T}} \\
    U_i^{\mathrm{T}}
  \end{pmatrix}
  \det (D-\lambda\mathbf{I}) \det
  \begin{pmatrix}
    (D-\lambda\mathbf{I})^{-1}\mathbf{u}_i &U_j
  \end{pmatrix},
\end{equation}
we rearrange the first matrix on the RHS utilizing the fact that a
swap of rows changes the sign of determinant
\begin{align}
  \det \left[ (B-\lambda \mathbf{I})^{(ij)} \right]
  &= (-1)^{i+j} \det (D-\lambda\mathbf{I}) \det
  \begin{pmatrix}
    \mathbf{u}_j^{\mathrm{T}} \\
    U_j^{\mathrm{T}}
  \end{pmatrix}
  \det
  \begin{pmatrix}
    (D-\lambda\mathbf{I})^{-1}\mathbf{u}_i &U_j
  \end{pmatrix} \nonumber \\
  &= (-1)^{i+j} \det (D-\lambda\mathbf{I})\det
  \begin{pmatrix}
    \mathbf{u}_j^{\mathrm{T}} (D-\lambda\mathbf{I})^{-1}\mathbf{u}_i & 0 \\
    U_j^{\mathrm{T}}(D-\lambda\mathbf{I})^{-1}\mathbf{u}_i & \mathbf{I}
  \end{pmatrix} \nonumber \\
  &= (-1)^{i+j} \det (D-\lambda\mathbf{I})
  \left[\mathbf{u}_j^{\mathrm{T}}
    (D-\lambda\mathbf{I})^{-1}\mathbf{u}_i\right] \; .
  \label{eq:det-of-minor}
\end{align}
We find this expression more agreeable than the expression we began with, as $
D-\lambda \mathbf{I} $ is a diagonal matrix.

From here on the proof of Eq. \eqref{eq:det11} is straightforward. First we note that
\begin{equation}
  \left. \dfrac{\mathrm{d}^n}{\mathrm{d} \lambda^n}
    (D-\lambda\mathbf{I})^{-1}  \right|_{\lambda=0} = n! \, D^{-(n+1)}
  \; ,
\end{equation}
furthermore, the determinant is simply a polynomial in $ \lambda $
\begin{equation}
  \det (D-\lambda\mathbf{I})
  = p_{N}\lambda^{N} + \ldots + p_1\lambda + p_0 \; ,
\end{equation}
from which we find that
\begin{equation}
  \left. \dfrac{\mathrm{d}^{n}}{\mathrm{d} \lambda^n}
    \det (D-\lambda\mathbf{I}) \right|_{\lambda=0} = n! \,p_{n} \; .
\end{equation}
The coefficients $ p_k $ are found from the product $\prod_i(D_{ii}-1)$. To optimize the calculation we introduce $ p_l^{(r)} $, which is the
coefficient obtained through the multiplication of
the first $ r $ factors in $\prod_i(D_{ii}-1)$. Thus, e.g., $ p_0^{(1)} = D_{11} $ and $ p_1^{(1)}=-1
$. The computation of $p_k$ is then performed iteratively 
\begin{equation}
  p^{(r+1)}_{l+1} = d_{r+1}p^{(r)}_{l+1}-p^{(r)}_l.
\end{equation}
This procedure at $r=N$ gives the coefficients of the
polynomial arising from the determinant. The computation is done in $O(N^2) $ steps. 
With this all parts of the final expression for the derivatives of
the determinants are ready. As advertised it is
\begin{align}
  \left[\dfrac{\partial^{l}}{\partial\lambda^{l}}
    \det\left[(B(x)-\lambda \mathbf{I})^{(ij)}\right]\right]_{\lambda=0}
  &= (-1)^{i+j} \sum_{n=0}^{l}\binom{l}{n} (l-n)!\,p_{l-n}\, n!\,
  \mathbf{u}_j^{\mathrm{T}} D^{-(n+1)} \mathbf{u}_i \nonumber \\
  &=(-1)^{i+j} l!\,\mathbf{u}_j^{\mathrm{T}} \left(\sum_{n=0}^{l}p_{l-n}
    D^{-(n+1)}\right) \mathbf{u}_i \; .
\end{align}

Finally, let us justify the assumption that none of the eigenvalues of $
B(x) $ is zero, because the entire procedure hinges on this
fact. Assume for contradiction that there exists a non-zero vector $
\mathbf{v} $ so that $ B\mathbf{v} = \mathbf{0} $, i.e.,
\begin{equation}
  \sum_{j=1}^{N} [B(x)]_{ij}v_j
  = \int_{a}^{x} \mathrm{d} y \, \psi_i(y)
  \sum_{j=1}^{N}v_j\psi_j(y)=0, \qquad \forall i\leq N.
\end{equation}
This holds for all $ i $, therefore, it must hold for linear
combinations too. Hence,
\begin{equation}
  0 = \sum_{i=1}^{N}v_i \int_{a}^{x}\mathrm{d} y \, \psi_i(y)
  \sum_{j=1}^{N}v_j\psi_j(y) = \int_{a}^{x}\mathrm{d} y
  \left(\sum_{i=1}^{N}v_i \psi_i(y)\right)^2 \Rightarrow \sum_{i=1}^{N}v_i \psi_i(y) = 0, \quad y\in [a,x].
  \label{eq:linear_combination}
\end{equation}
As $\mathbf{v}$ is non-zero, it should contain at least one
non-zero entry. Without loss of generality, let us assume $v_1 \neq 0$
(we can always redefine the index $i$ such that $v_1 \neq 0$). The
wave functions individually solve the one-body Schr{\"o}dinger
equation, therefore
\begin{align}
  \label{eq:zero_operator}
  \prod\limits_{k=2}^N (H_0 - E_k) \sum\limits_{i=1}^N v_i \psi_i(y)
  &= \sum\limits_{i=1}^N (E_i - E_2) (E_i - E_3) \cdots (E_i - E_N) v_i
  \psi_i(y) \nonumber\\
  &= (E_1 - E_2) (E_1 - E_3) \cdots (E_1 - E_N) v_1 \psi_1(y)
  = 0 , \quad y \in [a, x]
\end{align}
It is known that a one-body system in a one-dimensional trap without singularities has a
non-degenerate spectrum, i.e., all the $E_i$'s are distinct, therefore
 the factor $(E_1 - E_2) \cdots (E_1 - E_N)$ does not vanish. By
assumption $v_1 \neq 0$, thus,  $\psi_1(y)$ must vanish on
$[a, x]$. Now consider the differential equation
\begin{equation}
  (H_0 - E_1) \psi_1(y) = 0 , \quad y \in [a, b]
\end{equation}
which is fulfilled because $(H_0 - E_1)$ acts as the zero operator on
the entire interval $[a, b]$. The same differential equation is solved
by the zero function on $[a, b]$. Now as the solutions $\psi_1(y) $
and $ 0 $ (the zero function) coincide on the subinterval $ [a,x] $ it
follows from the theory of ordinary differential equations that given
sufficiently smooth $ V $, the two functions must be identical on the
whole interval $ [a,b] $. We are forced to conclude that
\begin{equation}
  \psi_1(y) = 0, \quad y\in[a,b],
\end{equation}
which is obviously not true, because by assumption, $\psi_1$ is  a non-zero eigenfunction of
$H_0$ on $[a,b]$. Thus, we arrive at the
contradiction, and must put $\mathbf{v} = \mathbf{0}$. This
proves that the $B(x)$-matrix does not have 
zero-eigenvalues that can cause problems in our algorithm.

%\section{Proof of Eq.~\eqref{eq:B_expanded}}
\section{Proof of Eq.~(14)}
\label{app:B_expanded}
\noindent Using equations~\eqref{eq:B-matrix} and
\eqref{eq:expansion-coefficient} we write the elements of the
$B(x)$-matrix as
\begin{align*}
  [B(x)]_{i,j} &
  = \sum_{n=1}^{N_b} \sum_{m=1}^{N_b} C_{i,m} C_{j,n} \frac{2}{L} \int_0^x \textrm{d}z
  \sin \left(\frac{m \pi z}{L}\right) \sin \left(\frac{n \pi z}{L}\right) \\
  &=\sum_{n=1}^{N_b} \sum_{m=1}^{N_b} C_{i,m} C_{j,n} \frac{1}{L} \int_0^x \textrm{d}z
  \left[
  \cos \left((m-n)\frac{\pi z}{L} \right)-\cos \left((m+n)\frac{\pi
      z}{L} \right) \right] \; .
\end{align*}
The integral yields $L [f(x)]_{m,n}$, where the function $f(x)$ is
defined in Eq.~\eqref{eq:f-matrix}, therefore,
\begin{align*}
  [B(x)]_{i,j} &= \sum_{n=1}^{N_b} \sum_{m=1}^{N_b} C_{i,m} [f(x)]_{m,n} C_{j,n} \\
  &= (Cf(x)C^\textrm{T})_{i,j} \; ,
\end{align*}
which is what we wanted to show.

\section{Scaling of the coefficients}
\label{sec:Appendix_Units_Scaling}
In this section we prove the scaling of coefficients in the case of
homogeneous potentials \eqref{eq:Scaling}.  For this we consider a
homogeneous potential, i.e., there exist a point $a \in (0,L)$ and a
real number $s$ such that for any $k \in \mathbb{R}$ the following
holds
\begin{equation*}
  V(k(x-a)) = k^s V(x-a).
\end{equation*}
Without loss of generality, we assume that $a=0$. This leads to the
dimensionless potential
\begin{align*}
  \tilde{V}(\tilde{x})
  = 2 \ell^2 \ell^s V(\tilde{x})=2 \ell^{s+2} V(\tilde{x})
\end{align*}
We see that if we scale the potential and $\ell$ at the same
time as follows
\begin{align*}
  V(x) \rightarrow \gamma V(x), \qquad \ell \rightarrow \ell'=\ell \gamma^{-\frac{1}{s+2}},
\end{align*}
the dimensionless potential will not change, and, therefore, the
dimensionless geometric coefficients returned by CONAN should be the
same, i.e.,
\begin{equation*}
\tilde{\alpha}_k \left[2 \ell^2 V(x/\ell) \right]
= \tilde{\alpha}_k \left[2\ell'^2\gamma V(x /\ell') \right],
\end{equation*}
which using Eq. \eqref{eq:coefficient_units} leads to the scaling
presented in Eq. \eqref{eq:Scaling}
\begin{align*}
  \alpha_k [\gamma V]
  = \gamma^{\frac{3}{s+2}} \alpha_k [V].
\end{align*}

\twocolumngrid

%%%%%%%%%%%%%%%%%%%%%%%%% BIBLIOGRAPHY %%%%%%%%%%%%%%%%%%%%%%%%%%%

\end{document}